\documentclass{IEEEtran}
\usepackage{bbm}
\usepackage{amssymb,amsmath,amsthm}
\usepackage{graphicx,epstopdf,psfrag,url,cite,xcolor,multirow,float}
\usepackage{booktabs}
\usepackage{caption-patch}
\usepackage[small,bf]{caption}
\usepackage{multirow}
\usepackage{makecell}

\usepackage[font=footnotesize]{subfig}
\usepackage[margin=0.7in]{geometry}

\usepackage{algorithm}
\usepackage{algorithmic}
\usepackage[super]{nth}
\usepackage{array}
\usepackage{pifont}
\usepackage{subfig}
\newcolumntype{C}[1]{>{\centering\let\newline\\\arraybackslash\hspace{0pt}}m{#1}}

\newtheoremstyle{mythm}% name
{\topsep}   % Space above1
{\topsep}   % Space below1
{\itshape}      % Body font
{0pt}       % Indent amount2
{\bfseries} % Theorem head font
{:}         % Punctuation after theorem head
{5pt plus 1pt minus 1pt}    % Space after theorem head3
{\thmname{#1}\thmnumber{ #2}\thmnote{ (#3)}}
\theoremstyle{mythm}

\newtheorem{proposition}{Proposition}

\newcommand{\bm}[1]{\boldsymbol{#1}}

\IEEEoverridecommandlockouts

\begin{document}

\title{AoI-aware Sensing Scheduling and Trajectory Optimization for Multi-UAV-assisted Wireless  Backscatter Networks}

\author{Yusi Long\IEEEauthorrefmark{1},
        Songhan Zhao\IEEEauthorrefmark{1},
        Shimin Gong\IEEEauthorrefmark{1},
        Bo Gu\IEEEauthorrefmark{1},
        Dusit Niyato\IEEEauthorrefmark{3},
        Xuemin (Sherman) Shen\IEEEauthorrefmark{4}
        \\
\IEEEauthorblockA{
\IEEEauthorrefmark{1}School of Intelligent Systems Engineering, Sun Yat-sen University, China\\
%\IEEEauthorrefmark{2}Guangdong Provincial Key Laboratory of Fire Science and Intelligent Emergency Technology, China\\
\IEEEauthorrefmark{3}Nanyang Technological University, School of Computer Engineering, Singapore\\
\IEEEauthorrefmark{4}Department of Electrical and Computer Engineering, University of Waterloo, Canada
}
}

\maketitle
\thispagestyle{empty}

\begin{abstract}
This paper considers multiple unmanned aerial vehicles (UAVs) to assist sensing data transmissions from the ground users (GUs) to a remote base station (BS).
Each UAV collects sensing data from the GUs and then forwards the sensing data to the remote BS. The GUs first backscatter their data to the UAVs and then all UAVs forward data to the BS by the non-orthogonal multiple access (NOMA) transmissions. We formulate a multi-stage stochastic optimization problem to minimize the long-term time-averaged age-of-information~(AoI) by jointly optimizing the GUs' access control, the UAVs' beamforming, and trajectory planning strategies. To solve this problem, we first model the dynamics of the GUs' AoI statuses by virtual queueing systems, and then propose the AoI-aware sensing scheduling and trajectory optimization (AoI-STO) algorithm. This allows us to transform the multi-stage AoI minimization problem into a series of per-slot control problems by using the Lyapunov optimization framework. In each time slot, the GUs' access control, the UAVs' beamforming, and mobility control strategies are updated by using the block coordinate descent (BCD) method according to the instant GUs' AoI statuses. Simulation results reveal that the proposed AoI-STO algorithm can reduce the overall AoI by more than 50\%. The GUs' scheduling fairness is also improved greatly by adapting the GUs' access control compared with typical baseline schemes.
\end{abstract}
\section{Introduction}
Recently,  the unmanned aerial vehicles (UAVs) have attracted extensive attention and been actively investigated in wireless networks.
Due to the significant advantages of fast mobility, flexible deployment, and enhanced line-of-sight (LoS) links between the UAVs and the ground users (GUs)~\cite{UAVsurvey},\cite{2021-Los-Seid}, the UAVs can be used to extend the network coverage when a large number of GUs are remotely distributed.
Considering the UAVs' hardware and resource constraints, e.g., limited onboard energy supply~\cite{2021UAV-LimitedEnergy}, computation and caching capabilities~\cite{2018UAV-mec-sherman}, the cooperative UAVs working together become more effective to improve the system performance in large-scale wireless networks, in terms of the system throughput~\cite{2022multi-uav-throughput-1,2022multi-uav-throughput-2, 2018shermanA-G-throughput}, the coverage extension~\cite{2022multiUAV-coverage_1,2018multiUAV-coverage_2, 2018sherman-UAV-deployment}, and data freshness at the information requesters~\cite{2022multiUAV-aoi-1,2022multiUAV-aoi-2,2022YusiUAV-AoI}, etc.
The flexibility and agility of the UAV-assisted wireless networks can help realize various real-time sensing applications that require quick responses to the GUs' sensing data, such as the UAV-assisted air quality monitoring~\cite{2018UAVair} and disaster rescue~\cite{2019Disaster-Management}.
In particular, the UAV-assisted system can provide temporary communication infrastructure.
It enables the rapid establishment of communication networks in damaged areas, promoting communication and coordination in various disaster and rescue scenarios.
The sensing data can be collected timely and processed effectively for real-time processing and decision making, especially when the sensing data changes rapidly. The obsolete information may lead to erroneous control, even disasters.
%Thus, the primary design problem is to optimize the transmission control that ensures the information freshness at the base station (BS) or back-end server.

Due to the UAV's fast mobility and on-board processing capability, the UAV-assisted wireless sensing networks can be reshaped dynamically to meet the GUs' time-varying traffic demands and quality provisioning requirements, calling for joint optimization of the UAVs' transmission control and trajectory planning to improve the information freshness at the base station (BS).
The information freshness is typically characterized by the age-of-information (AoI), defined as the time elapsed since the most recent data update event, which can be viewed as an evaluation on the value of the collected data from the perspective of the information requester~\cite{2017Aoi-origin}\cite{2023-AoI-Seid}.
It is expected that the GUs' AoI values at the BS can be reduced when the UAVs collect the GUs' sensing data more frequently and also forward the data to the BS efficiently with minimum delay~\cite{2020uav-sense-trans}.
AoI minimization in UAV-assisted networks is closely related to the data sensing and forwarding processes. Given the UAVs' locations, the GUs' access control can be optimized based on the GUs' data traffics and AoI statuses. The GUs' channel conditions can be controlled by the UAVs' beamforming to adapt the uplink rates of sensing data transmission. Considering the GUs' time-varying traffic demands, the UAVs' trajectories should be jointly optimized with the transmission control to balance all GUs' AoI performance.

One of the most salient challenges for the UAVs' sensing data collection lies in the channel competition among different GUs with limited energy supply.
%Typically, the numerous GUs can be wireless sensors and mobile smart devices, constituting a part of the future Internet of Things (IoT).
%Considering the GUs' limited energy supply,
To minimize the GUs' energy consumption, passive backscatter communication can be adopted for the GUs' uplink data transmissions by avoiding the use of power-consuming RF transceivers at the GUs.
As such, each UAV can be used as a mobile carrier emitter and the access point for the backscattering GUs.
Via trajectory planning, the UAVs can fly closer to energy-limited GUs on-demand to enhance their uplink transmission rates.
The GUs' access control is also required to optimize to minimize the overall AoI in the UAV-assisted wireless networks~\cite{2018multiUAV-tra-wu}.
This can avoid uncoordinated competition for uplink data transmissions to the UAVs, which may result in channel congestion and excessive transmission delay.
A simple intuition for the GUs' access control is that the GUs with higher waiting delays may have higher priorities to access the uplink transmission channels.
However, the optimal design of the GUs' access control is a non-trivial task in a complex network, especially with limited energy supply and channel resources.
%In particular, the GUs' access control is closely related to the UAVs' beamforming strategies.
%{\color{blue}The UAVs can achieve higher sensing rates with a cluster of GUs closed to each other by aligning with the GUs' channels, while the UAVs' beamforming optimization  becomes inefficient to maximize the separated GUs' uplink offloading rates simultaneously due to the different channel conditions.}
%When the GUs are close to each other and thus have a similar channel condition, the UAVs' beamforming strategies can be optimized to align with the GUs' channels, and thus the UAVs can achieve higher sensing rates. When the GUs are distant from each other with very different channel conditions, the UAVs' beamforming optimization becomes inefficient to maximize all GUs' uplink transmission rates simultaneously.
%Due to the combinatorial nature, it is challenging to design {\color{blue}a joint strategy that includes the GUs' access control, the UAVs' mobility control and beamforming} to reduce the overall AoI. This becomes more difficult when the GUs are subject to limited energy supply.
In particular, if a GU uploads its sensing data to the UAVs frequently, it will run out of energy faster and become inactive, while the less frequently scheduled GUs will experience the AoI's deterioration. After collecting the GUs' sensing data, the UAVs will forward the sensing data to the remote BS with the minimum transmission delay. In a spectrum sharing environment, the UAVs have to compete for the forwarding channels for sensing data transmissions. The transmission rates are also coupled with the UAVs' trajectory planning strategies. %Although the conventional time-division multiple access protocol can avoid channel contention by assigning each UAV a dedicated time slot, it may induce excessive transmission delay and limit the channel utilization.
%The above challenges motivate our work in this paper to construct a more efficient {\color{blue}joint control} strategy to minimize the overall AoI.
All the above challenges motivate our work in this paper to construct an efficient joint control strategy for the GUs' access control, the UAVs' beamforming, and mobility control strategies to minimize the overall AoI of all GUs in the UAV-assisted wireless networks.

In this paper, the design objective is to minimize the long-term time-averaged  AoI in a multi-UAV-assisted wireless network, which comprises a single-antenna BS, multiple multi-antenna UAVs, and a large number of single-antenna GUs. The UAVs can efficiently enhance wireless connectivity and expand coverage by dynamically adapting their trajectories. Specifically, the UAVs first hover over specific sensing locations and provide carrier signals for the GUs when performing data sensing via low-power backscatter communications. Subsequently, they forward the sensing data to the BS and then move to the next sensing locations.
The joint optimization of the GUs' access control, the UAVs' beamforming, and mobility control strategies is spatial-temporally coupled in different time slots, leading to a high-dimensional mix-integer dynamic program that is difficult to solve practically.
The stochastic AoI minimization problem is first decomposed by Lyapunov optimization framework into a series of per-slot control problems. Then, the main task focuses on the joint optimization of the GUs' access control, the UAVs' trajectory planning, and beamforming optimization in each time slot.
In particular, the per-slot control problem mainly aims to optimize the data sensing and forwarding capacities of the multi-UAV-assisted wireless network.
The UAVs' sensing capacities rely on the optimal planning for the UAVs' flying, sensing, and forwarding phases in a time-slotted frame structure. A longer flying time implies that the UAVs are expected to find better positions for data sensing and forwarding. A longer sensing time means that the UAVs can collect more data from the GUs, while a longer forwarding time ensures successful data transmission to the remote BS.
In the UAVs' data forwarding phase, the resolution of the UAVs' channel competition is required in a spectrum-sharing environment. The conventional time-division protocol can be problematic and resource-inefficient to coordinate the UAVs in a dynamic wireless network due to the UAVs' fast mobility.
In this paper, the non-orthogonal multiple access (NOMA) is considered as a more spectrum-efficient alternative to support the UAVs' data forwarding by allowing multiple UAVs with different channel conditions to transmit in the same channel simultaneously. The information from different UAVs can be decoded sequentially by using superposition coding at the UAVs and successive interference cancellation at the BS~\cite{2022UAV-NOMA-Adam}. The capacity improvement can be significant when the UAVs are spatially separated with very different channel conditions.
It is expected that the UAVs can transmit the collected sensing data more efficiently by using the NOMA transmissions, and thus reduce the overall AoI at the BS.
Specifically, the main contributions of this paper are summarized as follows:
\begin{itemize}
\item \emph{UAV-assisted backscatter sensing and NOMA transmissions:}
In the UAV-assisted sensing phase, the UAVs emit carrier signals to low-power GUs to support uplink sensing data transmissions through backscattering.
Given the UAVs' deployment locations, the UAVs' beamforming strategies are further optimized by reshaping the UAV-GU channel conditions to help coordinate real-time sensing data transmissions from multiple GUs in dynamic and complex network environments.
This prevents data from becoming outdated due to unpreferable channel conditions and scheduling delay.
In the UAV-assisted forwarding phase, the UAVs' channel diversities are exploited due to their rapid mobility and spatial separation. The NOMA is used to improve spectrum efficiency by allowing different UAVs to forward their sensing data simultaneously.
The NOMA transmissions along with the UAVs' trajectory planning are expected to  exploit and even create orthogonal channel opportunities to improve the spectrum efficiency.
%This implies that multiple UAVs can simultaneously transmit sensing data without scheduling delay, ensuring the quick data delivery to the BS.}
%The low-backscatter communications and NOMA transmissions are employed in the UAVs' sensing and forwarding phases respectively to improve the sensing and transmission capacity. In the sensing phase, the GUs backscatter the sensing data to the UAVs. Via active beamforming control, the multi-antenna UAVs not only improve the GUs' uplink transmission rates but also adapt the GUs' access control to maximize the sensing efficiency. In the UAV-assisted forwarding phase, we exploit the UAVs' channel diversity due to fast mobility and spatial separation, and rely on NOMA to improve spectrum efficiency by allowing different UAVs to forward their sensing data simultaneously.
\item \emph{Per-slot decomposition via Lyapunov optimization:} The long-term AoI minimization can be formulated by jointly optimizing the GUs' access control, the UAVs' trajectory planning, and the UAVs' beamforming strategies. The AoI dynamics are firstly modeled as virtual queues. Then the stochastic long-term AoI minimization is decomposed into a sequence of per-slot control subproblems by using the Lyapunov optimization framework. The control subproblem in each time slot involves the GUs' access control, the UAVs' trajectory planning, and beamforming strategies, given the GUs' AoI statuses and data traffics.
\item \emph{AoI-aware sensing scheduling and trajectory optimization (AoI-STO):} After decomposition, the AoI-STO algorithm is proposed to solve the per-slot control subproblem in each time slot by iterative three steps.
    The first step is to adapt the GUs' access control strategy to maximize each UAV's sensing capacity. This can be achieved by reshaping the UAV-GU channel conditions when the UAVs' locations and the beamforming strategies are fixed.
    The second step is to optimize the UAVs' beamforming strategies to maximize the UAVs' sensing and transmissiona. This can improve the UAVs' relay capacities by strengthening the UAVs' signals.
    The third step is to update the UAVs' trajectories by optimizing the UAVs' hovering locations and the time allocation for flying, sensing, and forwarding phases.
    Extensive simulation results reveal that the AoI-STO scheme can significantly reduce the overall AoI compared with the baseline schemes.
    It also improves the GUs' scheduling fairness by adapting the GUs' access control strategy.
\end{itemize}

Some preliminary results of this work have been reported in \cite{2022YusiUAV-AoI}. This paper further extends the study in \cite{2022YusiUAV-AoI} by proposing backscatter-aided data sensing and multi-UAV-assisted NOMA transmissions to reduce the overall AoI by improving the transmission capacity. This paper incorporates multiple antennas on each UAV to further explore the performance gain between the UAVs' beamforming and the GUs' access control. More extensive simulation results are also provided to verify that the UAVs' beamforming and trajectories along with the GUs' access control can balance the GUs' virtual AoI queue and minimize the overall AoI.
The remainder of the paper is organized as follows.  A literature review is presented in Section \ref{sec:work}. The system model is introduced in Section \ref{sec:model}.
The AoI minimization problem is formulated and decomposed by Lyapunov optimization in Section \ref{sec:lya}. The per-slot control problem is determined and its solution is presented in Section \ref{sec:per}. Section \ref{sec:eva} presents the simulation results and Section \ref{sec:con} concludes the paper.

\section{Related work}\label{sec:work}
\subsection{UAV-assisted Wireless Sensing Networks}
AoI in the UAV-assisted wireless networks is closely related to the UAVs' trajectory planning and the GUs' access control strategies.
When a UAV is far away from the GUs or more specifically the UAV-GU channel condition becomes worse off, a longer sensing time is required to collect the sensing data from the GUs.
If all UAVs are deployed closer to the GUs, this will result in strong conflicts among them and poor forwarding performance to the remote BS, which brings up the overall AoI.
Therefore, it is of great importance for the UAVs to adjust their trajectories to improve the sensing capacities.
The authors in~\cite{2022UAVsensingZhu} aimed to maintain data freshness by optimizing the UAV's hovering position with a path search algorithm.
The authors in~\cite{2019sherman-UAB-3DTra} proposed to reduce the UAV-GU path-loss by optimizing a group of UAVs' three-dimensional (3D) trajectories and the GUs' scheduling strategy in UAV-assisted wireless networks.
The authors in~\cite{2022UAVsensingXiong} proposed to enhance the secure and energy-efficient data collection under an eavesdropping attack by planning the UAVs' trajectories.
Given the UAVs' hovering positions, some GUs may have inferior channel conditions, which limit the uplink transmission rates.
In this case, the GUs' access control or sensing scheduling is required to improve the sensing efficiency, e.g., the GUs' data transmissions can be postponed till their channel conditions become much better as the UAVs move to more preferable positions.
The authors in~\cite{2021-UAV-aoi-choudhury} achieved a superior AoI performance by scheduling the GUs according to their channel information caused by the change of the UAVs' trajectories.
The authors in~\cite{2023-UAV-AoI-wang} proposed a learning approach to jointly optimize the UAVs' trajectories and the GU's scheduling to reduce the system AoI. Different from the single-GU access control in~\cite{2021-UAV-aoi-choudhury} and~\cite{2023-UAV-AoI-wang}, more flexible access control strategies can be explored in complex wireless networks with various resource constraints. %The authors in~\cite{2022wang-scheduling-aoi} achieved a better performance gain by selecting a group of GUs with a larger AoI and a higher probability of successful transmission.
The authors in~\cite{2020-UAV-multiuser-scheduling} studied a multi-GU access control strategy that allows a group of GUs to upload their data to the central controller with NOMA method. The real-time access control decision was determined by collecting time-varying GUs' preferences and diverse QoS requirements.
The authors in~\cite{2023-UAV-MADDPG} aimed to use adaptive learning method to minimize the long-term average task completion delay by adapting real-time multi-GU access control.
The authors in~\cite{2020-UAV-multiuser-zeng} revealed that it is beneficial for the UAVs to interact with the environment frequently due to the fluctuating interference. Such information can help develop an agile multi-GU access control strategy to improve the system sum rate.
Different from the previous works, it is revealed that the GUs' access control depends on both the UAVs' beamforming strategies and the sensing locations in this paper.
The GUs' access control generally relies on the energy status, channel and the AoI conditions.
All these conditions are related to the UAVs' beamforming and trajectory strategies.
However, their joint optimization has been seldom studied in the literature.

\subsection{Backscatter-aided AoI Minimization}
The low-power backscatter communications allow the GUs to upload sensing data when the GUs' energy supply becomes insufficient for RF communications. The authors in~\cite{2021UAV-backscatter-aoi} focused on a backscatter-aided scenario where a single-antenna UAV collects data from multiple GUs one by one and finally carries the data to the BS. The use of backscatter communications is to save the GUs' power consumption for uplink data transmissions. The average AoI of all GUs is minimized by jointly optimizing the UAV's data collection time and trajectory. The authors in~\cite{2022UAVaoi-cons} employed backscatter communications in cognitive radio networks when the primary user occupies the licensed spectrum with high probability. To fight against the dynamic environment, a deep learning method was used to ensure data timeliness by learning the time and energy allocation. Instead of AoI minimization, the authors in~\cite{2021backscatter-delay} focused on the overall transmission delay in backscatter-aided and wireless-powered mobile edge computing systems. The overall delay for data offloading and computation is minimized by jointly optimizing the operation time of a power beacon, the computing frequency, transmit power, and portions of workload for backscatter offloading. Moreover, the reconfigurability of passive intelligent reflecting surface (IRS) can be used to enhance the channel quality and keep the information fresh. The authors in~\cite{2022IRS-aoi} proposed to reduce the transmission delay by controlling the BS's beamforming vector and the IRS's phase shifting matrices. The IRS can be also used to mitigate signal propagation impairments from the UAV to the GUs and thus ensure fresh data collection by jointly optimizing the UAV's trajectory, the GU's scheduling, and the IRS's discrete phase shifts~\cite{2022tvt-uav-irs-aoi}.
An aerial IRS (AIRS) can be carried by the UAV and used to improve the LoS links between the IRS and the GUs. The authors in~\cite{2022-AIRS} exploited the AIRS to improve the information freshness while fulfilling the real-time user demands. The successive convex approximation (SCA) algorithm was proposed to minimize the sum AoI by jointly optimizing the active and passive beamforming strategies, as well as the UAV's trajectory and transmission scheduling strategies.
The authors in~\cite{2021UAVIRS-AOI} used the AIRS as a relay from the GUs to the BS, which demonstrates superior AoI degeneration.
In this paper, backscatter communications are similarly utilized to assist in uploading sensing information to the UAVs. Different from the above works, the UAVs' sensing-transmission tradeoff is the key to minimize the overall AoI in this paper. Note that the time allocation for data sensing and forwarding is confined in one time slot. More sensing time can be allocated to the GUs only if the UAVs' forwarding capacities are improved via NOMA transmissions. Hence, the backscatter-aided sensing time and the UAV-assisted transmission time should be jointly optimized and adapted according to the UAVs' trajectory planning and beamforming strategies.

\subsection{Multi-UAV NOMA Transmissions}
Once the sensing data arrives at the UAVs, it is preferable for the UAVs to  forward them to BS with the minimum delay.
The primary task is to fulfill this design target by maximizing the transmission capacity from the UAVs to the BS. To improve the network throughput, the authors in~\cite{2022channelUAV} proposed the frequency-division technique to collect sensing data simultaneously from multiple GUs in a UAV-assisted post-disaster network. The authors in~\cite{2021uavTDMA} employed the time-division scheme to allocate each GU a sensing slot. The GU's sensing scheduling and the UAV's trajectory are jointly optimized to maximize the system throughput and expand the communication coverage. Recently, the NOMA technique was shown to have significant performance improvements in large-scale wireless networks, such as high spectral efficiency, massive connectivity, and low latency.
The authors in~\cite{2018UAVNOMA-So} considered a UAV-assisted multi-user wireless system, in which the UAV is used as a flying BS. The UAV's high mobility can provide the enhanced UAV-GU connection to serve the GUs with NOMA transmissions. The network coverage and the system throughput can be enhanced by jointly optimizing the UAV's altitude, power allocation, and bandwidth allocation.
The authors in~\cite{2022UAVPowerAllocation} focused on a multi-UAV-assisted vehicular communication network and aimed to maximize the system capacity by using the NOMA transmissions. The authors in~\cite{2022-UAV-NOMA-CR} studied a multi-UAV-assisted cognitive radio network, where the secondary users transmit data to each UAV simultaneously and the UAVs deliver data to the destination. The throughput of the secondary network can be significantly improved by optimizing the UAV's power allocation in NOMA transmissions. The authors in~\cite{2022UAV-NOMA-WBAN} focused on a UAV-assisted large-scale wireless body area network for remote monitoring of the patients' vital signs by optimizing each UAV's trajectory. The NOMA technique was employed to simultaneously schedule UAVs' data transmissions, which can enhance the network throughput with high spectrum efficiency.
Different from the above work, the main task is to utilize the UAVs' NOMA transmissions to reduce the overall AoI in this paper. To improve the sensing capacity, the UAVs can be deployed to extend the sensing coverage by allocating each UAV a different service region. This may limit the UAVs' forwarding capacity due to a larger distance between the UAVs and the remote BS.
By using NOMA transmissions, the UAVs' performance loss can be effectively compensated in the forwarding phase. However, the performance analysis of NOMA transmissions is complicated by the UAVs' trajectory planning and beamforming strategies.

\begin{table*}[ht]
\caption{A Summary of UAV-assisted wireless sensing networks}
\label{table}
\centering
\resizebox{13.3cm}{!}{
\begin{tabular}{|c|c|c|c|c|} %±í¸ñ4ÁÐ È«²¿¾ÓÖÐÏÔÊ¾
\hline
\textbf{UAV-assisted} & \textbf{{UAV-GU Link (UL/DL)}} & \textbf{ Performance Metric} & \textbf{Control Variables} & \textbf{Algorithm Design} \\
\hline
Single UAV & UL RF commu. &UAV's trajectory & AoI& ML \cite{2022UAVsensingZhu} \\
\hline
Single UAV & UL RF commu. &UAV's trajectory, band multi-user scheduling & Sum rate
& AO \cite{2020-UAV-multiuser-scheduling} \\
\hline
Single UAV & UL RF commu. &UAV's trajectory, multi-user scheduling and power allocation
& Sum rate & AO \cite{2020-UAV-multiuser-zeng} \\
\hline
Single UAV & UL backscatter commu. &UAV's data collection time and trajectory & AoI & Heuristic algorithm \cite{2021UAV-backscatter-aoi} \\
\hline
Single UAV & UL IRS commu. &\makecell[c]{UAV's trajectory, the GU's scheduling, \\and the IRS's discrete phase shifts}
 & AoI & DRL \cite{2022tvt-uav-irs-aoi} \\
\hline
Single UAV & AIRS reflection &\makecell[c]{Active and passive beamforming strategies, \\the UAV's trajectory and transmission scheduling strategies} & AoI & SCA \cite{2022-AIRS} \\
\hline
Single UAV & AIRS reflection &Altitude of the UAV and phases-shift of RIS & AoI & DRL \cite{2021UAVIRS-AOI} \\
\hline
Single UAV & UL RF TDMA commu. &UAV's  trajectory and single user scheduling & Sum rate & AO \cite{2021uavTDMA} \\
\hline
Single UAV & DL RF NOMA commu. &UAV's altitude, power allocation, and bandwidth allocation & Sum rate & AO \cite{2018UAVNOMA-So} \\
\hline
Single UAV & UL RF NOMA commu. &UAV's trajectory and single user scheduling & Sum rate & Q-learning \cite{2022UAV-NOMA-WBAN} \\
\hline
Multiple UAVs & DL RF FDMA commu. &Channel allocation and single user scheduling & Sum rate & Stackelberg game \cite{2022channelUAV}  \\
\hline
Multiple UAVs & UL RF commu. &UAVs'  trajectories and single user scheduling & Channel gain & AO \cite{2019sherman-UAB-3DTra} \\
\hline
Multiple UAVs & UL RF commu. &UAVs'  trajectories and single user scheduling & Secure sum rate & AO, SCA \cite{2022UAVsensingXiong}  \\
\hline
Multiple UAVs & UL RF commu. &Single user scheduling & AoI & Heuristic algorithm, DQN \cite{2021-UAV-aoi-choudhury}  \\
\hline
Multiple UAVs & UL RF commu. &UAVs' trajectories and single user scheduling & AoI & DRL \cite{2023-UAV-AoI-wang}  \\
\hline
Multiple UAVs & DL RF NOMA commu. &UAVs' deployment design, and resource allocation of vehicles & Sum rate & Stackelberg Game \cite{2022UAVPowerAllocation}   \\
\hline
\end{tabular}}
\end{table*}

\begin{figure}[t]
    \centering
	\subfloat[A multi-UAV-assisted wireless sensing network]{
		\includegraphics[width=0.45\textwidth]{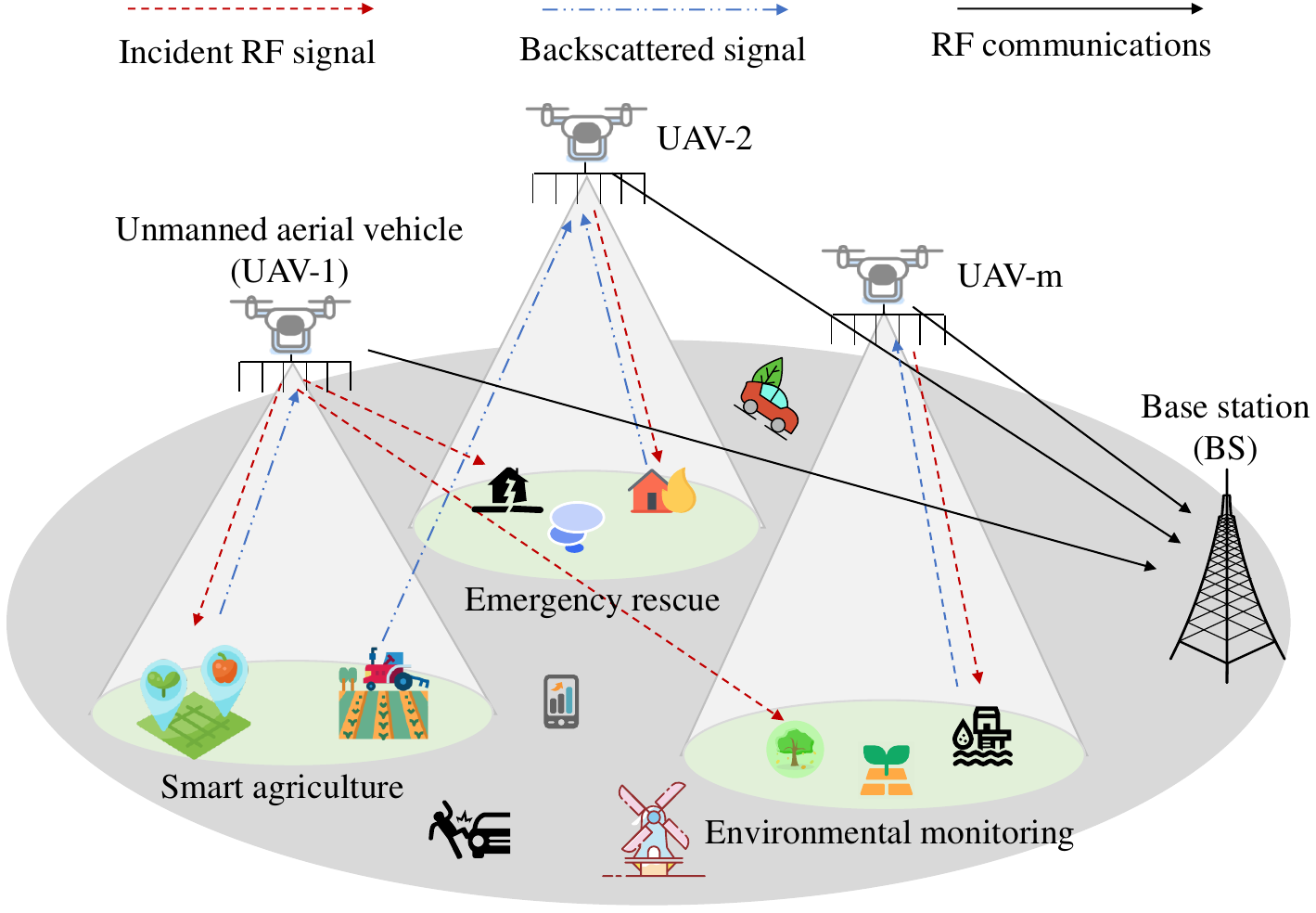}} \\
	\subfloat[Time allocation for the UAVs' sensing, flying, and forwarding phases]{
		\includegraphics[width=0.45\textwidth]{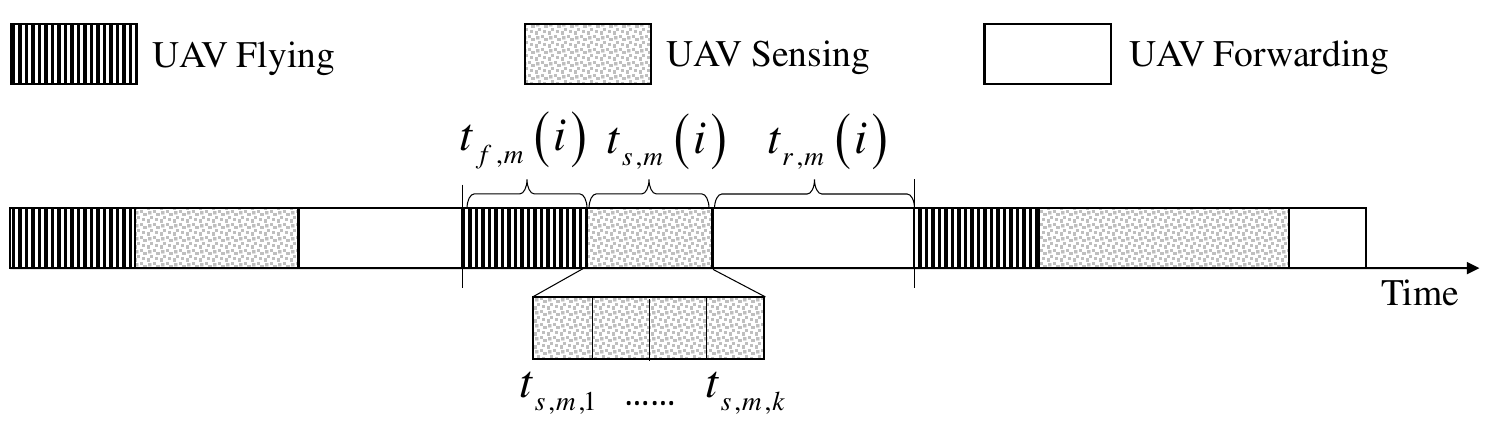}}
	\caption{The UAVs' planning for sensing, flying, and forwarding phases in a multi-UAV-assisted wireless network.}\label{fig_system_model}
\vspace{-0.4cm}
\end{figure}
\section{System model}\label{sec:model}
The multi-UAV-assisted wireless network consists of one BS, $K$ GUs and $M$ UAVs.
The sets of the GUs and the UAVs are denoted as $\mathcal{K}\triangleq\{1,2,...,K\}$ and $\mathcal{M}\triangleq\{1,2,...,M\}$, respectively.
Each UAV is equipped with $N$ antennas.
It is assumed that the single-antenna GUs cannot be served by the BS directly due to the blockage of surrounding obstacles. The UAVs are deployed to collect sensing data from $K$ GUs, randomly distributed in an open area, and then forward the collected sensing data to the BS for information update, as illustrated in Fig.~\ref{fig_system_model}(a). Similar to~\cite{2021UAV-backscatter}, each GU equipped with a passive backscatter communication module is capable of transmitting its sensing data to the UAV by backscattering the incident RF signal from the UAV.
A time-slotted multi-access protocol is considered as shown in Fig.~\ref{fig_system_model}(b). Each frame is divided into multiple time slots with a unit length. The set of all time slots is denoted by $\mathcal{I}\triangleq\{1,2,...,T\}$.

To keep the sensing data fresh at the BS, the UAVs need to optimize their data sensing and forwarding strategies jointly to minimize the sensing and transmission delays.
In this paper, the 3D coordinate is considered, where the locations of the UAV-$m$ and the GU-$k$ in the $i$-th time slot are denoted as $\bm{\ell}_m(i)=(x_m(i), y_m(i),z_m(i))$ and ${\bm q}_k = (x_k, y_k,0)$, respectively. Without loss of generality, all UAVs fly at a fixed altitude, i.e., $z_m(i) = H$. However, the following problem formulation and solution methods below can be easily extended to 3D trajectory planning.
The location of the BS's receiver antenna is denoted as ${\bm q}_0 = (x_0, y_0,z_0)$, which can be viewed as the GU-0 for notational convenience. The distance between the UAV-$m$ and the GU-$k$ is expressed as $d_{s,m,k}(i) = \|\bm \ell_m(i)-\bm q_{k}\|$ for $k \in \mathcal{\tilde{K}}\triangleq \mathcal{K} \cup\{0\}$ and $m \in \mathcal{M}$, and the distance between the UAV-$m$ and any UAV-$\tilde{m}$ is denoted as $d_{u,m,m'}(i) = \|\bm \ell_m(i)-\bm \ell_{m'}(i)\|$ for $m' \in \mathcal{M}$.
%\begin{align}\label{equ:dis-UG}
%d_{m,\tilde{k}}(i) = \sqrt{\|\bm \ell_m(i)-\bm q_{\tilde{k}}\|^2},
%\end{align}
\subsection{Fly-to-Sense-and-Forward (FSF) Protocol} \label{subsec:UAV}
Each FSF time slot is further divided into three sub-slots $\bm{t}_m(i)=\left(t_{f,m}(i), t_{s,m}(i), t_{r,m}(i)\right)$ for the UAVs' flying, sensing, and forwarding phases, as shown in Fig.~\ref{fig_system_model} (b). Thus, the feasible time allocation is as follows:
\begin{align}\label{con:time-fea}
t_{s,m}(i)\! + \!t_{f,m}(i) \!+\! t_{r,m}(i)\leq 1,\quad \forall m \in \mathcal{M} \text{ and } i\in\mathcal{T}.
\end{align}
In the sub-slot $t_{f,m}(i)$, each UAV flies to a hovering position and then receives up-to-date sensing data from the GUs under its signal coverage in the sub-slot $t_{s,m}(i)$. In the sub-slot $t_{r,m}(i)$, the UAVs forward the sensing data to the BS by the NOMA transmissions. The UAVs may not perform data sensing (or forwarding) when they are far away from the GUs (or the BS). According to the GUs' traffic demands, the UAVs' time allocation for flying, sensing, and forwarding needs to be optimized to ensure information freshness at the BS.
%\begin{figure}[t]
%	\centering	\includegraphics[width=0.45\textwidth]{fig2-time.pdf}
%	\caption{Time-slotted structure for UAV sensing, flying and reporting}\label{fig_time}
%\end{figure}

%In each time slot, a larger flying duration implies that each UAV can not only cover more GUs for information sensing, but also fly closer to the BS for data reporting, {\color{blue}a larger sensing period means that the UAVs can receive more data traffic}, and a larger reporting time implies that each UAV can provide a higher transmission throughput to BS and a better quality of service for communication.

%\subsubsection{Flexible flying of UAVs}
 %The purpose of each UAV's flight is to find the suitable location that not only enables more GUs to upload their data to associated UAVs, but also encourages the UAVs to generate greater transmission throughput to complete data reporting.
 %For any different $ m, m' \in \mathcal{M}, m\neq m'$,
 %In practice,
The UAVs' trajectories need to avoid collision and be subject to the speed limit $v_{\max}$ in each time slot $i \in \mathcal{I}$, i.e.,
\begin{subequations}\label{con:trajectory}
\begin{align}
&d_{u,m,m'}(i) \geq d_{\rm{min}}, \quad \forall m, m' \in \mathcal{M} \text{ and } m\neq m',  \label{con:collision}\\
&\|\bm{\ell}_m(i)-\bm{\ell}_m(i-1)\| \leq t_{f,m}(i)v_{\max},\quad \forall m \in \mathcal{M}, \label{con:speed}
\end{align}
\end{subequations}
where $d_{\rm{min}}$ is the minimum distance between two UAVs to ensure safety, and $v_{\max}$ denotes the UAVs' maximum flying speed~\cite{2018multiUAV-tra-wu}. The channel conditions depend on the distance between the transceivers. For any $m \in \mathcal{M}$ and $k \in \mathcal{\tilde{K}}$, let $\mathbf{h}_{m,k}(i)$ denote the complex UAV-GU channel vector:
\begin{align}\label{equ:com-channel-UG}
 \mathbf{h}_{m,k}(i) \!\! =\! \! \sqrt{\!\!\rho d_{s,m,k}^{-2}\!(i) }  \!\!\left(\! \sqrt{\!\frac{g_0}{g_0 \!+ \!1}}\mathbf{\bar{g}}_{m,k}\!(i) \!\!+\!\! \sqrt{\!\frac{1}{g_0 \!+ \!1}}\mathbf{\tilde{g}}_{m,k} \!(i)\!\! \right)\!\!,
\end{align}
where $\rho$ represents the channel power gain at the reference distance of 1 meter. The Rician factor $g_0$ combines the LoS and the Rayleigh fading components, denoted as $\mathbf{\bar{g}}_{m,k}(i)$ and $\mathbf{\tilde{g}}_{m,k}(i)$, respectively~\cite{Zhang19TWC_UAV}.
\subsection{Backscatter-aided Data Sensing}
The sensing process involves the GUs' data generation, access control, and uplink transmissions.
The generation of each GU's sensing data is assumed to be independently and identically distributed.
%Let $\varsigma_k(i)=1$ be the indicator that new sensing data is generated at GU-$k$ in time slot $t$ and $\varsigma_k(i)=0$ otherwise.
It is assumed that the sensing data has a very small size, which represents the status change in the sensing environment.
In each time slot, the sensing data is randomly generated at each GU-$k$ and stays in its data queue until it is collected by the UAV or replaced by the newly generated sensing data.
%Each GU has opportunity to access the corresponding UAV to backscatter the sensing information.
Given the UAVs' hovering positions, multiple GUs can upload their sensing data via backscatter communications in a time-division protocol.

A binary variable $\beta_{m,k}(i)$ is defined to denote the GUs' access control.
The $\beta_{m,k}(i)$ is set to 1 if the GU-$k$ is associated with the UAV-$m$ in the $i$-th time slot, and $\beta_{m,k}(i)=0$ otherwise.
In the $i$-th time slot, each UAV can serve multiple GUs with the time-division protocol while each GU can be associated with at most one UAV.
Hence, the GUs' access control constraints can be expressed as follows:
\begin{align}
\!\!\!\!\sum_{m=1}^{M}\beta_{m,k}(i)\leq 1, \beta_{m,k}(i)\in \{0,1\}, \forall m \in \mathcal{M}  \text{ and } k \in \mathcal{K}.\label{con:association-UG-a}
 \end{align}
%Each UAV collects information from the associated GUs one by one in the sensing sub-slot $t_{s,m}(i)$.
Let $t_{s,m,k}$  denote the constant mini-slot for the GU-$k$ to upload its data to the UAV-$m$. The UAV-$m$'s sensing scheduling is also limited by the total sensing time, i.e.,
\begin{align}\label{equ:sen-time}
t_{s,m}(i) \geq \sum_{k=1}^{K}\beta_{m,k}(i)t_{s,m,k}, \quad \forall  m \in \mathcal{M}.
\end{align}
\vspace{-0.3cm}
%Specifically, in the sensing period, each UAV emits carrier signals to activate the GUs {\color{blue}via RF beamforming} and simultaneously {\color{blue}receives the backscattered information from the GUs.} Then, we denote the set of these scheduled GUs, which corresponds to different UAVs but operates at the same time, as $\mathcal{K}_c$.
%Moreover, the UAVs perform data sensing in a cooperative manner. Specifically,
%When UAV-$m$ collects the sensing data from GU-$k$, it will provide carrier signals to other GUs to assist other UAVs improving sensing data rate by enhancing the communication channels.
%Besides, when the UAV-$m$ is interacting with BGU-$k$, it will be interfered by other BGU-$k' \in \mathcal{K}_c, k' \neq k$.

It is assumed that each UAV-$m$ has two sets of antennas, i.e., one for sensing beamforming and the other for data  reception. Let $\mathbf{w}_{s,m}(i) \in \mathbb{C}^{N\times 1}$ and $\mathbf{w}_{r,m}(i) \in \mathbb{C}^{N\times 1}$ denote the normalized transmit and receive beamforming vectors of the UAV-$m$ in the sensing sub-slot $t_{s,m}(i)$, respectively. The sensing beamforming vector $\mathbf {w}_{s,m}(i)$ is used to control the direction and strength of the carrier signals for the GUs' backscatter communications. The UAV-$m$'s beamforming signal is given by $\mathbf{u}_m(i) = \sqrt{p_s}\mathbf{w}_{s,m}(i) x_{s,m}$, where $p_s$ denotes the transmit power and $x_{s,m}$ is a random information symbol with unit power. Thus, the incident signal at the GU-$k$ can be formulated as $c_k(i)=\sum_{m \in \mathcal{M}}\mathbf{h}_{m,k}^H(i)\mathbf{u}_m(i)$. Meanwhile, the GU-$k$ modulates its sensing data on the incident carrier signal $c_k(i)$ by controlling the reflection coefficient $\Gamma(i)=\Gamma_o x_{u,k}(i)$, where $\Gamma_o$ is an antenna-specific constant coefficient and $x_{u,k}(i)$ is the backscattered information symbol with unit power. Note that each GU-$k$ works in the time-division protocol and thus the receive beamforming can be aligned to the UAV-$m$'s channel vector, i.e., $\mathbf{w}_{r,m}(i)=\mathbf{h}_{m,k}(i)/\|\mathbf{h}_{m,k}(i)\|.$ This can maximize the GU-$k$'s uplink transmission rate.
%the signal received by the UAV-$m$ is given by $\mathbf{y}_{m,k}(i) = x_{s,m}\Gamma_o \mathbf{h}_{m,k}^{H}(i)x_{u,k}(i)c_k(i)\!+\!\mathbf{v}_u$,
%{\color{blue}\begin{align}\label{equ:rec-signal-UAV}
%\!\!\!\mathbf{y}_{m,k}(i) &= \mathbf{w}_{sr,m}(i) \left(x_{s,m}(\Gamma_o \mathbf{h}_{m,k}^{H}(i)x_{u,k}(i)c_k(i)\!+\!\mathbf{v}_u \right),
%\end{align}}
%where $\mathbf{v}_u$ denotes the noise signals at each multi-antenna UAV and can be normalized to the unit power for simplicity.
%By using the MRC scheme, the received signal strength at the UAV-$m$ from GU-$k$ is given by $\gamma_{m,k}(i)=|\Gamma_o|^2\|\mathbf{h}_{m,k}(i)\mathbf{h}^H_{m,k}(i)\mathbf{w}_{s,m}(i)\|^2 = |\Gamma_o|^2\rho^2 \mathbf{Tr}\left(\mathbf{G}_{m,k}(i) \mathbf{G}^H_{m,k}(i)\mathbf{W}_{s,m}(i)\right)/\|\bm \ell_m(i)-\mathbf{q}_k\|^4$. Here we introduce the matrix variables $\mathbf{G}^H_{m,k}(i)\triangleq \mathbf{g}_{m,k}(i)\mathbf{g}^H_{m,k}(i)$ and $\mathbf{W}_{s,m}(i)\triangleq \mathbf{w}_{s,m}(i)\mathbf{w}^H_{s,m}(i)$ to simplify the representation of $\gamma_{m,k}(i)$,  following the
%semi-definite relaxation.
%Without loss of generality, assuming $\mathbb{E}\left[|x_{u,k}(i)|^2\right]=1$.
%the received signal strength at the UAV-$m$ from BGU-$k$ is $\gamma_{m,k}(i) = p_s|\Gamma_o|^2 \|\mathbf{h}_{m,k}(i)\|^2 \sum\limits_{\tilde{m} \in \mathcal{M}} |\mathbf{h}_{\tilde{m},k}^{H}(i)\mathbf{w}_{s,\tilde{m}}(i)|^2$.
As such, the received signal-to-noise-ratio (SNR) of the GU-$k$ at the UAV-$m$ can be represented as follows:
\begin{align}\label{equ:SINR-UG}
\gamma_{m,k}(i) =p_s|\Gamma_o|^2 \left|\mathbf{h}_{m,k}^{H}(i)\right|^2 \sum_{m'=1 }^{M} \left|\mathbf{h}_{m',k}^{H}(i)\mathbf{w}_{s,m'}(i)\right|^2.
\end{align}
%$\gamma_{m,k}(i) \!\!=\!\! p_s|\Gamma_o|^2 \|\mathbf{h}_{m,k}^{H}(i)\|^2 \!\!\sum\limits_{\tilde{m} \in \mathcal{M}} \!\!|\mathbf{h}_{\tilde{m},k}^{H}(i)\mathbf{w}_{st,\tilde{m}}(i)|^2$.
%\begin{align}\label{equ:SINR-UG}
%\!\!\!\!\gamma_{m,k}(i) \!\!=\!\! p_s|\Gamma_o|^2 \!|\mathbf{h}_{m,k}^{H}(i)|^2 \!\!\!\sum\limits_{\tilde{m} \in \mathcal{M}} \!\!\!|\mathbf{h}_{\tilde{m},k}^{H}(i)\mathbf{w}_{s,\tilde{m}}(i)|^2 \!, k \in \mathcal{K}.
%\end{align}
%\vspace{-0.5em}
By collecting all sensing data from the active GUs under the UAV-$m$'s coverage, the UAV-$m$'s sensing capacity  in the $i$-th time slot is denoted as follows:
\begin{align}\label{equ:ori-sm}
s_{m}(i) = \sum_{k=1}^{K}\beta_{m,k}(i)t_{s,m,k}\log_2 \Big(1+\gamma_{m,k}(i)\Big).
\end{align}
%{\color{blue}Note that} the sensing data for status update is of a small size.
The GU can effectively transmit the sensing information to the UAV in a constant mini-slot $t_{s,m,k}$.
%\begin{align}\label{equ:sensing-data-traffic}
%s_{s,m}(i) = \sum_{k \in \mathcal{K}}r_{s,m,k}(i)t_{s,m,k}.
%\end{align}
 %The UAV's data queue is limited by the maximum capacity $Q_{\max}$. The consumed energy of UAV-$m$ when collecting sensing data from GUs is denoted by:
 %\begin{align}\label{equ:UAV-sensing-erergy}
 %e_m(i) = \mathbf{Tr}(\mathbf{W}_m(i))\sum_{k \in \mathcal{K}}\alpha_{m,k}(i)t_{s,m,k}(i).
 %\end{align}
%Meanwhile, the peak power constraint at the UAV-$m$ when collecting sensing data is as follow:
%\begin{align}\label{equ:UAV-sensing-peak-power}
%0\leq \mathbf{w}_{s,\tilde{m}}(i)\leq 1,
%\end{align}
%where $p_{s,\max}$ is the maximum transmit power of the UAV when collecting sensing data.
%The GUs associated with the UAV can transmit information by backscattering the incident signals from the UAV, and other GUs can harvest RF energy from the UAV. Thus, the RF energy harvested by the idle GU-$k$ in one time period is given by:
%\begin{align}\label{equ:har-energy-GU}
%E_k(i) = \sum_{\tau=1}^{t}\sum_{m=1}^{M}\frac{\left(1-\alpha_{m,k}(\tau)\right)\rho \mathbf{Tr}(\mathbf{G}_{m,k}(i)\mathbf{W}_m(i))}{\|\bm \ell_m(i)-\mathbf{q}_k\|^2} t_{s,m}(i)
%\end{align}
\subsection{Multi-UAV NOMA Transmissions}
When the UAVs complete data sensing, they hover in the air and forward the collected data to the BS via the NOMA transmissions. Let $\mathbf{w}_{t,m}(i) \in \mathcal{C}^{N\times 1}$ denote the UAV-$m$'s transmit beamforming vector in the forwarding sub-slot $t_{r,m}(i)$.
To reduce waiting delay, the forwarded information from the UAVs with less sensing data can be decoded first at the BS. This can ensure that the UAV with a larger data size has a higher transmission rate. Without abuse of notations, let $\mathcal{M}$ denote the ordered set of UAVs according to their data sizes such that $0 < s_{1}(i)\leq s_{2}(i)\leq...\leq s_{M}(i)$. Thus, the BS firstly decodes the UAV-$1$'s signal and considers all other UAVs' signals as the interference. After the UAV-$1$'s decoding, its signal is subtracted from the received signals and then the UAV-$2$'s signal is subsequently decoded. The decoding procedure continues until the signal of the last UAV is successfully received by the BS. The received signal at the BS from the UAV-$m$ is denoted by $p_s|\mathbf{h}_{m,0}^H(i)\mathbf{w}_{t,m}(i)|^2$. Define $\gamma_{m,0}(i)=\sum_{m'=m}^{M}p_s|\mathbf{h}_{m',0}^H(i)\mathbf{w}_{t,m'}(i)|^2$ as the interference power for simplicity. Thus, the received throughput from the UAV-$m$ to the BS can be denoted as follows:
\begin{align}\label{equ:ori-dm}
r_{m}(i)= t_{r,m}(i) \log_2 \left(\frac{1+\gamma_{m,0}(i) }{1+\gamma_{{m+1},0}(i)}\right), \quad \forall m \in \mathcal{M}.
\end{align}
Here a normalized noise power is considered for simplicity.

\section{Lyapunov Optimization for AoI Minimization}\label{sec:lya}
%\subsection{Data-based Age of Information} \label{subsec:AOI}
Let $a_k(i)$ be the GU-$k$'s AoI at the beginning of the $i$-th time slot.
The successful delivery of sensing data from one GU to the BS depends on two conditions: $1)$ the GU successfully uploads its data to a UAV, and $2)$ the UAV successfully forwards the sensing data to the BS, i.e.,
\begin{align}\label{equ:sm-dm}
s_{m}(i)\leq r_m(i), \quad \forall m \in \mathcal{M}.
\end{align}
If the GU-$k$ is not allowed to access any UAV in the $i$-th time slot, its sensing data will be further delayed by one time slot, and thus its AoI is updated as $a_k(i+1) = a_k(i) + 1$.
If the GU-$k$ uploads its data to the UAV-$m$ and the UAV-$m$ successfully forwards the sensing data to the BS, its AoI in the next time slot will decrease.
Furthermore, if all sensing data within the UAV-$m$'s coverage can be forwarded to the BS successfully in the $i$-th time slot, the GUs' AoI values will drop to zero in the next time slot, i.e., $a_k(i+1) = 0$.
\subsection{AoI Dynamics and Long-term AoI Minimization}

\begin{figure}[t]
	\centering \includegraphics[width=0.45\textwidth]{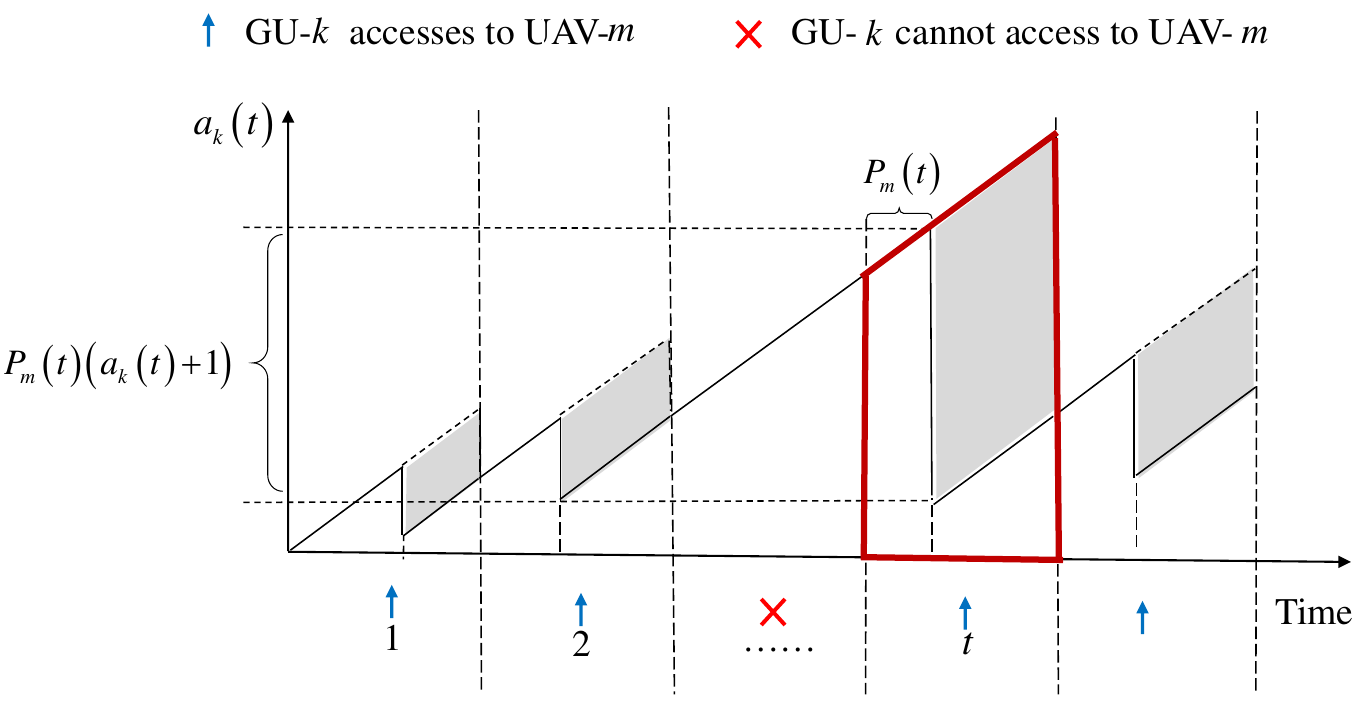}
	\caption{The AoI dynamics of the GU-$k$.}\label{fig_AoIstru}\vspace{-0.7cm}
\end{figure}

Different from complete data transmissions in~\cite{R4}, the UAV-$m$ may not be able to collect all sensing data successfully from the GUs within the sensing sub-slot $t_{s,m}(i)$ due to limited channel capacity. In this case, let $P_m(i)\triangleq \frac{s_{m}(i)}{s_{c,m}(i)}$ denote the fraction of successfully collected data by the UAV-$m$, where $s_{c,m}(i)$ is the size of sensing data within the UAV-$m$'s coverage before data collection and $s_m(i)$ is the sensing capacity defined in Equation~\eqref{equ:ori-sm}.
By this definition, the GUs' AoI can be updated partially if only a part of the sensing data is successfully forwarded to the BS. Hence, with $\beta_{m,k}(i)=1$, the GU-$k$'s average AoI can be correspondingly updated as $(1-P_m(i))(a_k(i)+1)$.
As such, the GUs' AoI dynamics can be summarized as follows:
\begin{align}
a_k(i+1)=
\begin{cases}
(1-P_m(i))(a_k(i)+1), & \beta_{m,k}(i)=1, \label{equ:aoi-ori}\\
a_k(i)+1, & \text{otherwise}.
\end{cases}
\end{align}
An illustrative example of the AoI dynamics is shown in Fig.~\ref{fig_AoIstru}.
For notational convenience, the AoI dynamics in~\ref{equ:aoi-ori} can be reformulated in a compact form as follows:
\begin{align} \label{equ:aoi}
a_k(i+1) =\left(1 -\sum_{m=1}^{M}\beta_{m,k}(i)P_m(i)\right)\Big(a_k(i)+1 \Big).
\end{align}
The binary access control $\beta_{m,k}(i)$ allows the GU-$k$ to connect with at most one UAV in each time slot, as shown in~\eqref{con:association-UG-a}.

Given the AoI's limit $a_{\max}$, the time-averaged AoI of each GU is upper bounded as follows:
\begin{equation}\label{con:age}
\lim_{T \rightarrow \infty} \frac{1}{T} \sum_{i=0}^{T-1} \mathbb{E}\big[a_k(i+1)\big] \leq a_{\max}, \quad \forall k \in \mathcal{K}.
\end{equation}
The expectation is taken with respect to the GUs' access control strategy.
The design target is to minimize the long-term time-averaged AoI of all GUs by optimizing the GUs' access control $\Phi\triangleq\{\beta_{m,k}(i)\}_{ {m\in\mathcal{M}, k\in\mathcal{K}, i \in \mathcal{T}}}$, the UAVs' mobility $(\bm \ell, \bm t)\triangleq({\bm\ell}_m(i), {\bm t}_m(i))_{m \in \mathcal{M}, i \in \mathcal{T}}$, and beamforming strategies $(\mathbf{w}_{s}, \mathbf{w}_{t})\triangleq(\mathbf{w}_{s,m}(i), \mathbf{w}_{t,m}(i))_{m \in \mathcal{M}, i \in \mathcal{T}}$ in data sensing and forwarding phases. The feasible region of the beamforming strategies $(\mathbf{w}_{s}, \mathbf{w}_{t})$ is given as follows:
\begin{align}\label{con:beamforming}
\{\|\mathbf{w}_{s,m}(i)\|\leq 1, \|\mathbf{w}_{t,m}(i)\|\leq 1,   \forall m \in\mathcal{M}, \forall i \in\mathcal{T}\}.
\end{align}
It is clear that the AoI performance has complicated couplings with the above control variables. For simplicity, the time-averaged AoI is defined as follows:
\begin{align}\label{equ:averageAoI}
\!\!\bar{A}(\Phi, {\bm \ell}, {\bm t}, {\mathbf{w}_{s}}, {\mathbf{w}_{t}}) = \lim_{T\rightarrow\infty}
\frac{1}{TK}\mathbb{E}\left[\sum_{i=0}^{T-1}\sum_{k=1}^K a_k(i+1)\right].
\end{align}
Till this point, the AoI minimization problem is formulated as follows:
%\begin{subequations}
\begin{align}\label{prob:aoi-reduction}
\min_{\Phi, {\bm \ell}, {\bm t}, {\mathbf{w}_{s}}, {\mathbf{w}_{t}}}~\bar{A}(\Phi, {\bm \ell}, {\bm t}, {\mathbf{w}_{s}}, {\mathbf{w}_{t}}), \ \quad{\rm s.t.} ~\eqref{con:time-fea}-\eqref{con:beamforming}.
\end{align}
Problem in Equation~\eqref{prob:aoi-reduction} is challenging to solve due to the following reasons. Firstly, the optimization of the GUs' access control strategy is combinatorial as it defines a discrete feasible set. The UAVs' beamforming optimization also affects the GUs' access control, which leads to a high-dimensional mix-integer program. Secondly, even with the fixed access control strategy, the UAVs' trajectory planning and time allocation are spatial-temporally coupled in a stochastic form. A dynamic programming approach to solve this problem can be practically intractable due to the curse of dimensionality.

\subsection{Per-slot Decomposition via Lyapunov Optimization}
\begin{figure}[t]
	\centering \includegraphics[width=0.45\textwidth]{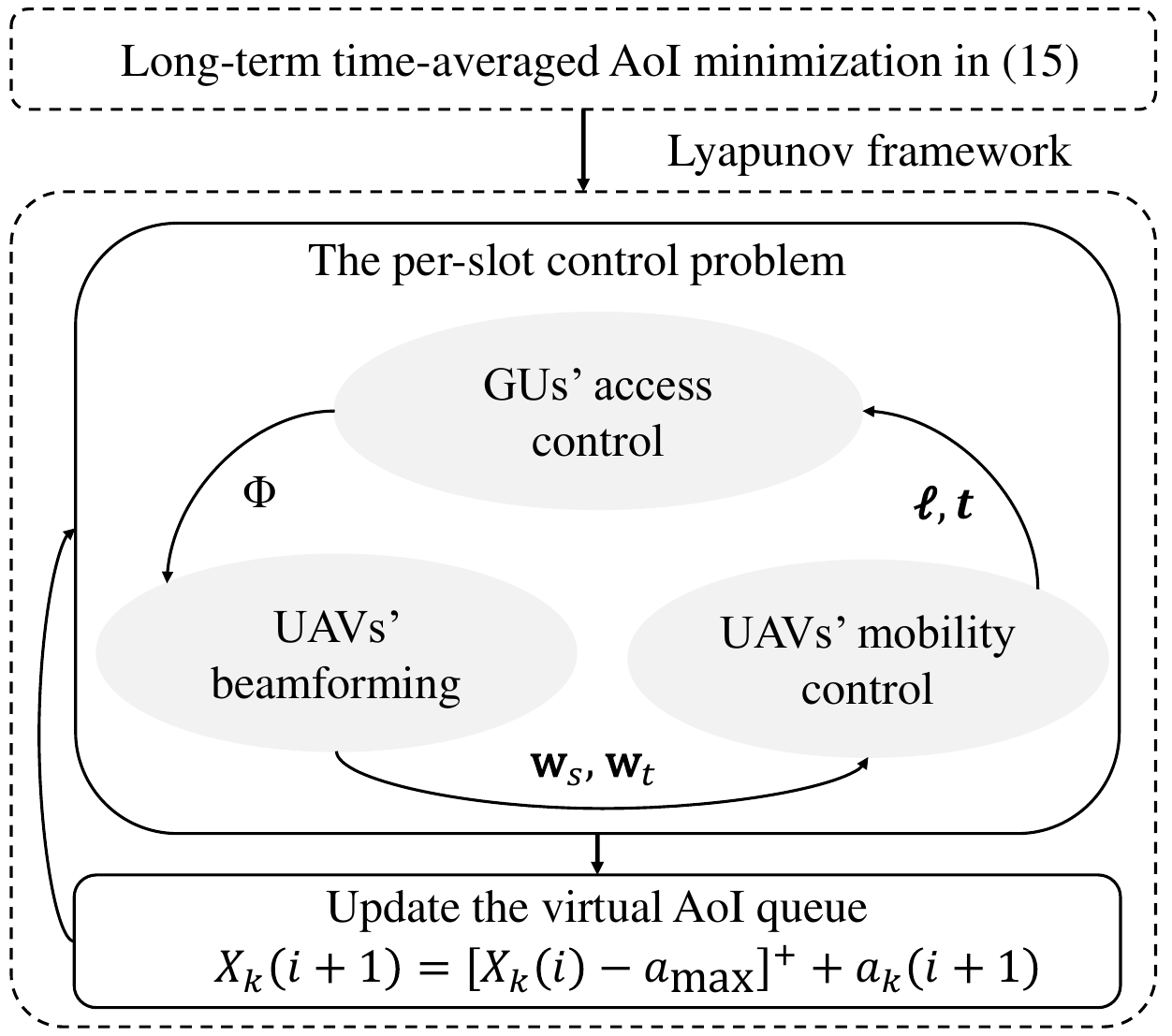}
	\caption{The overall algorithm framework }\label{fig_overall}\vspace{-0.7cm}
\end{figure}

The long-term time-averaged AoI minimization in Equation~\eqref{prob:aoi-reduction} is obviously a stochastic and dynamic program with high computational complexity.
It involves resource allocation in multiple time slots and varying system states over time, which is practically difficult to solve. The GUs' access control and the UAVs' mobility control in each time slot not only depend on the current system states but also affect the future evolution of the GUs' AoI statuses. To resolve this difficulty, the Lyapunov optimization framework is first employed to decompose the multi-stage stochastic AoI minimization problem into a series of per-slot control sub-problems in different time slots by introducing queue stability constraints, as illustrated in Fig.~\ref{fig_overall}.

To proceed, the Proposition~\ref{prop-queue} is first given as follows to show a simplified reformulation of the time-averaged constraint in~\eqref{con:age}. The reformulation stems from the conclusion in~\cite{inequality2queue}, which provides a generalized method to approximate a stochastic inequality by using a virtual queue system.
\begin{proposition}\label{prop-queue}
For each GU-$k$, $k \in \mathcal{K}$, a virtual queue $X_k(i)$ can be constructed with initial zero state, i.e., $X_k(0)=0$, and the queue dynamics given by:
\begin{equation}
X_k(i+1) =\big[X_k(i)-a_{\max}\big]^+ + a_k(i+1).\label{equ:AoI_queue}
\end{equation}
If $X_k(i)$ is mean rate stable, i.e., $\lim_{i\rightarrow\infty}\frac{1}{i}E[|X_k(i)|]=0$, the satisfaction of the inequality in \eqref{con:age} can be ensured.
\end{proposition}
The proof of Proposition~\ref{prop-queue} is relegated to Appendix A. Proposition \ref{prop-queue} implies that each GU has the bounded AoI value if the access control and the UAVs' mobility control strategies can ensure the stability of the virtual AoI queues, i.e., the averaged queue size approaches zero as time progresses.
Therefore, in the following part, the stochastic inequality constraint in~\eqref{con:age} can be replaced by the stability constraint of the virtual AoI queue, i.e., $\lim_{i\rightarrow\infty}\frac{1}{i}E[|X_k(i)|]=0$.
%{\color{blue}Thus, the virtual AoI queue can be a prerequisite for using the Lyapunov algorithm subsequently}.

Denote $\mathbf{X}(i) = (X_1(i),...,X_K(i))$ as the state vector of all GUs' virtual AoI queues. The stability of $\mathbf{X}(i)$ can be measured by introducing the following Lyapunov function:
\begin{align}\label{equ:Lyapunov function}
&~{L}\big(\mathbf{X}(i)\big) = \frac{1}{2} \sum_{k=1}^{K} |X_k(i)|^2, \quad \forall \,i \in \mathcal{I},
\end{align}
which is a non-negative quadratic form of the virtual AoI queue states. The constant help ease our deduction and algorithm design in the following part. Given the virtual AoI queue states, if the Lyapunov function has a small value, all GUs' virtual AoI queues have small state values. Otherwise the Lyapunov function becomes large if at least one GU's virtual AoI has a large state value and tends to be unstable.
Therefore, the queue stability can be further characterized by using the expected change of the Lyapunov function in successive time slots, which is termed as the drift of the Lyapunov function  and denoted as follows:
\begin{align}
\label{equ:Lyapunov Drift}
&~{\Delta}_L\big(\mathbf{X}(i)\big) = \mathbb{E}\big[{L}\big(\mathbf{X}(i+1)\big)-{L}\big(\mathbf{X}(i)\big)\big|\mathbf{X}(i)\big].
\end{align}
Given the current queue state $\mathbf{X}(i)$, the expectation in~\eqref{equ:Lyapunov Drift} is taken over all GUs' access control, the UAVs' beamforming, and trajectory planning strategies in the $i$-th time slot.

To stabilize the virtual AoI queue $\mathbf{X}(i)$, it is required to minimize the increment of the queue size, i.e., the Lyapunov drift ${\Delta}_L(\mathbf{X}(i))$. Meanwhile, minimizing all GUs' AoI values is necessary to keep information fresh.
Thus, the minimization objective in each time slot is as follows:
\begin{equation}\label{equ:drift-penalty}
T(\mathbf{X}(i))\triangleq{\Delta}_L(\mathbf{X}(i))+ V\sum_{k=1}^{K}\mathbb{E}\big[a_{k}(i+1)\big|\mathbf{X}(i)\big],
 \end{equation}
where the constant $V$ is a non-negative control parameter to balance each GU's AoI and the queue stability. To this point, the stochastic objective in  Equation~\eqref{prob:aoi-reduction} can be replaced by the new minimization target in~\eqref{equ:drift-penalty}, and thus the optimization problem becomes a per-slot control problem with the known states of all virtual AoI queues. However, it is still difficult to minimize~\eqref{equ:drift-penalty} directly. Instead, an upper bound to~\eqref{equ:drift-penalty} is derived and minimized as an approximation, as revealed in Proposition~\ref{prop-bound}.
\begin{proposition}\label{prop-bound}
$T(\mathbf{X}(i))$ in~\eqref{equ:drift-penalty} is upper bounded as follows:
\begin{align}
& T(\mathbf{X}(i)) \leq  B \label{equ:Lya_penalty}\\
&-\sum_{k=1}^{K} \sum_{m=1}^{M} \mathbb{E}\left[\beta_{m,k}(i)P_m(i)\Big(V+X_k(i) \Big)\Big( a_k(i)+ 1\Big)|\mathbf{X}(i)\right] ,\nonumber
\end{align}
where $B = \sum_{k=1}^{K} \Big[\frac{1}{2} \Big(a^2_k(i)+2(X_k(i)+V+1) a_k(i)+a_{\max}^2+2X_k(i)+2V+1\Big)-X_k(i)a_{\max}\Big]$ is a finite constant given the current virtual AoI queue state.
\end{proposition}

The derivation of Proposition \ref{prop-bound} is relegated to Appendix B.
Given the current queue state $X_k(i)$ and $a_k(i)$, the finite constant $B$ is known at the beginning of each time slot. The control variables in~\eqref{equ:Lya_penalty} include the GUs' access control, time allocation, the UAVs' trajectories, and beamforming variables. The expectation is taken with respect to all available queue states $\mathbf{X}(i)$.
Instead of minimizing the objective in~\eqref{equ:drift-penalty} directly,
the focus now becomes the minimization of the upper bound in~\eqref{equ:Lya_penalty}.
%we now focus on the minimization of the upper bound in {\color{blue}Equation}~\eqref{equ:Lya_penalty}.
For simplicity, the time index can be dropped and the finite constant $B$ in~\eqref{equ:Lya_penalty} can be ignored. Once the queue states are observed at the beginning of the $i$-th time slot, the minimization of $T(\mathbf{X}(i))$ in~\eqref{equ:drift-penalty} can be approximated by the following maximization problem:
\begin{subequations}\label{prob:AoI-lya}
\begin{align}
\max_{\Phi, {\bm \ell}, {\bm t}, {\mathbf{w}_{s}},{\mathbf{w}_{t}}} ~ &\sum_{m=1}^M\sum_{k=1}^{K}\beta_{m,k}P_m(V+X_k)(a_k+1) \label{obj:AoI-lya}\\
 {\rm s.t.}~~~~~&\eqref{con:time-fea}-\eqref{equ:sm-dm}~ \text{and}~\eqref{con:beamforming}.
% &\eta_m\leq s_{s,m}, \forall m \in \mathcal{M}, \label{con:aux-sm}\\
 %& \eta_{v_m}\leq d_{v_m}, \forall v_m \in \mathcal{M}_v,\label{con:aux-dm}\\
\end{align}
\end{subequations}
Referring to the AoI dynamics in~\eqref{equ:aoi}, the objective function in~\eqref{obj:AoI-lya} can be viewed as the summation of individual GU's AoI reduction, i.e.,~$\sum_{m=1}^M\beta_{m,k}P_m(a_k+1)$, weighted by the parameter $V+X_k$. A larger AoI reduction implies that the GU's AoI performance can be reduced  significantly once it transmits its information successfully to the BS. Given the current AoI queue status $\mathbf{X}(i)$, the constant parameter $V+X_k$ denotes the importance of the $k$-th GU to the overall AoI performance. Hence, problem~\eqref{prob:AoI-lya} aims to schedule the data transmissions of the GUs with the higher AoI reduction and the worse AoI values in the current time slot.

Problem~\eqref{prob:AoI-lya} is a mixed-integer problem, which is still difficult to solve optimally.
The following part details present an iterative algorithm to address the problem~\eqref{prob:AoI-lya} by the block coordinate descent (BCD) and the SCA techniques.
The overall algorithm sketch is shown in Fig.~\ref{fig_overall}. A heuristic method is first proposed to adapt the GUs' access control.
Secondly, given the GUs' access control and the UAVs' mobility control decisions, the UAVs' beamforming strategies can be optimized for efficient data collecting and forwarding. In the third step, the UAVs' hovering locations in the next time slot and the time allocation for the sensing, flying, and forwarding phases can be efficiently optimized. After the above three-step iteration, the GUs' virtual AoI queue states can be updated in the next time slot according to~\eqref{equ:AoI_queue}.

\section{Per-slot Access Control, Beamforming Optimization, and Mobility Control}\label{sec:per}
%\subsection{Heuristic algorithm for GUs' access control}
Given the GUs' virtual AoI queue states, the per-slot control problem can be decomposed into three sub-problems: the GUs' access control, the UAVs' beamforming optimization, and the mobility control sub-problems. The GUs' access control aims to reduce the overall AoI.
Our intuitive design is to select the GUs with higher AoI values to upload the sensing data to the UAVs.
%, {\color{blue}as shown in Fig.~\ref{fig-heri-algo}.
All GUs are first ordered by their current AoI values and then a group of GUs with larger AoI values are selected to upload their sensing data sequentially to the UAV-$m$.
Specifically, given the UAVs' time allocation and hovering positions, the first step is to evaluate the number of active GUs under each UAV's coverage.
Then, given the UAV-$m$'s sensing time $t_{s,m}$, the maximum number of GUs can be calculated by $\zeta_{c,m} = \frac{1-t_{f,m}^{(\tau)}-t_{r,m}^{(\tau)}}{t_{s,m,k}}$, where $t_{s,m,k}$ is the mini-slot allocated to each active GU.
When the GUs are densely deployed, it is necessary to select a total number of $\zeta_{c,m}$ GUs from all GUs under the UAV-$m$'s
coverage.
The above heuristic method for the GUs' access control provides an efficient solution with low complexity that is easy to implement in practice.
%\begin{figure}[t]
%	\centering
%	\includegraphics[width=0.43\textwidth]{fig-heri-alo.pdf}
%	\caption{{\color{blue}The specific optimization process of the GUs' access control}.}\label{fig-heri-algo}
%\vspace{-0.3cm}
%\end{figure}
\subsection{The UAVs' Beamforming Optimization}
Given the GUs' access control strategy and the UAVs' hovering locations, the UAVs will collect the GUs' sensing data and then forward it to the BS via the NOMA transmissions. The UAVs can control the beamforming strategies in both the sensing and forwarding phases. The sensing beamforming optimization has to balance the uplink transmission rates of different GUs, while the forwarding beamforming optimization aims to exploit the channel diversity and orthogonality to maximize the network capacity via the UAVs' NOMA transmissions.
To simplify the beamforming optimization problem, the slack variables $x_{m,k}\geq |\mathbf{h}_{m,k}^H\mathbf{w}_{s,m}|^2$ and $y_{m}\geq |\mathbf{h}_{m,0}^H\mathbf{w}_{t,m}|^2$ for $ m \in \mathcal{M}$ and  $k \in\mathcal{K}$ are first introduced. Thus, the received signal at the UAV-$m$ from the GU-$k$ can be reformulated as $\tilde{\gamma}_{m,k} = p_s|\Gamma_o|^2 |\mathbf{h}_{m,k}^{H}|^2 \sum_{m' =1}^{M} x_{m',k}$. Similarly, the received signal at the BS from the UAV-$m$ can be expressed by $\tilde{\gamma}_{m,0} = p_s\sum_{m'=m}^{M} y_{m'}$. Thus, the UAVs' sensing and transmission beamforming strategies can be optimized by solving the following subproblem:
\setlength\belowdisplayskip{7pt}
\begin{subequations}\label{subprob:beam}
\begin{align}
\max_{\mathbf{w}_{s,m}, \mathbf{w}_{t,m}} ~&\sum_{m=1}^{M}\sum_{k=1}^{K}\beta_{m,k}P_m(V+X_k)(a_k+1) \\
{\rm s.t.}~~~&x_{m,k}\!\geq\! |\mathbf{h}_{m,k}^H\mathbf{w}_{s,m}|^2, \,\, y_{m}\geq |\mathbf{h}_{m,0}^H\mathbf{w}_{t,m}|^2, \label{con:x-wsm}\\
%~&~y_{m}\geq |\mathbf{h}_{m,0}^H\mathbf{w}_{t,m}|^2, \label{con:y-wtm}\\
~&s_m\leq \sum_{k=1}^{K}t_{s,m,k}\log_2(1+\tilde{\gamma}_{m,k}),  \label{con:eta-wsm}\\
 %&\sum_{k \in \mathcal{K}}\!\!t_{s,m,k}\!\log_2(1\!+\!\tilde{\gamma}_{m,k})\! \!\leq \! t_{r,m}\! \log_2 \!\!\left(\!\!\frac{1+\tilde{\gamma}_{m,0} }{1\!+\!\tilde{\gamma}_{{m+1},0}}\!\right)\!, \label{con:sm-dm-xy}\\
~&r_m \geq \sum_{k=1}^{K}t_{s,m,k}\log_2(1+\tilde{\gamma}_{m,k}) , \label{con:sm-dm-xy}\\
~&\|\mathbf{w}_{s,m}\| \leq  1, \|\mathbf{w}_{t,m}\| \leq  1,\\
~& \forall \,m \in \mathcal{M}  \text{ and }  k \in \mathcal{K}.
\end{align}
\end{subequations}
Problem~\eqref{subprob:beam} only focuses on the beamforming strategies while assuming fixed values for all other control variables.

The SCA technique is further applied to approximate \eqref{con:sm-dm-xy} with a convex function.
Given the solutions $x_{m,k}^{(\tau)}$ and $y_m^{(\tau)}$ in the $\tau$-th iteration, the constraint \eqref{con:sm-dm-xy} in the $\tau+1$-th iteration  can be approximated as follows:
\begin{align}\label{con:sm-dm-convex}
\!\! t_{r,m}\Big( \log_2 \left(1+\tilde{\gamma}_{m,0}\right)-F(y_m)\Big) \geq \sum_{k =1}^{K}t_{s,m,k}E(x_{m,k}),
\end{align}
where $F(y_m)$ is the linear approximation of $\log_2 \left(1+\tilde{\gamma}_{m+1,0}\right)$ and similarly $E(x_{m,k})$ denotes the linear approximation of $\log_2(1+\tilde{\gamma}_{m,k})$ detailed as follows:
\begin{align}
& E(x_{m,k}) \triangleq \log_2 \left( 1+ \tilde{\gamma}_{m,k}^{(\tau)} \right) \label{equ:Ek} \\
&+ {p_s|\Gamma_o|^2 \Big|\mathbf{h}_{m,k}^{H}\Big|^2}\Big(1+\tilde{\gamma}_{m,k}^{(\tau)}\Big)^{-1}\sum\limits_{m' =1}^{M} \left( x_{m',k}-x_{m',k}^{(\tau)} \right).\nonumber
\end{align}
%Similarly, $F(y_m)$ is the linear approximation of $\log_2 \left(1+\tilde{\gamma}_{m+1,0}\right)$.
%\begin{align}
%F_{m,0}^{\rm{lb}}  \triangleq \log_2 \left(1+\tilde{\gamma}_{m+1,0}^{(\tau)}\right)
%+\frac{\tilde{\gamma}_{m+1,0}-\tilde{\gamma}_{m+1,0}^{(\tau)}}{1+\tilde{\gamma}_{m+1,0}^{(\tau)}}. \nonumber
%\end{align}
Another difficulty lies in the quadratic terms $|\mathbf{h}_{m,k}^{H}\mathbf{w}_{s,m}|^2$ and $|\mathbf{h}_{m,0}^H\mathbf{w}_{t,m}|^2$ in~\eqref{con:x-wsm}. The matrix variables such that $\mathbf{W}_{s,m} = \mathbf{w}_{s,m}\mathbf{w}_{s,m}^H$ and $\mathbf{W}_{t,m} = \mathbf{w}_{t,m}$$\mathbf{w}_{t,m}^H$ can be further introduced.
Then, the received signal at the UAV-$m$ from the GU-$k$ can be reformulated as $\hat{\gamma}_{m,k} = p_s|\Gamma_o|^2 |\mathbf{h}_{m,k}^{H}|^2 \sum_{m'=1}^{M}\mathbf{Tr}\left(\mathbf{h}_{m',k}^H\mathbf{W}_{s,m'}\mathbf{h}_{m',k}\right)$.
Consequently, the UAV-assisted beamforming strategies can be optimized in the $\tau$-th iteration as follows:
\begin{subequations}\label{subprob:beamforming}
\begin{align}\label{obj:beamdorming}
\max_{\mathbf{W}_{s,m},\mathbf{W}_{t,m}} ~&~ \sum_{m=1}^{M}\sum_{k=1}^{K}\beta_{m,k}P_m(V+X_k)(a_k+1) \\
 {\rm s.t.} ~&~ {\rm{Rank}}(\mathbf{W}_{j,m})=1,  \forall j  \in  \{s,t\}, \label{con:rank-w}\\
~&~\mathbf{Tr}(\mathbf{W}_{j,m}) \leq  1, \mathbf{W}_{j,m}  \succeq   0,   \forall j  \in  \{s,t\},  \label{con:matrix-w}\\
~&~x_{m,k}\!\geq \!\mathbf{Tr}\!\left(\mathbf{h}_{m,k}^H\!\mathbf{W}_{s,m}\mathbf{h}_{m,k}\!\right), \label{con:xm} \\
~&~ y_{m}\geq\mathbf{Tr}(\mathbf{h}_{m,0}^H\mathbf{W}_{t,m}\mathbf{h}_{m,0}), \label{con:ym}\\
~&~s_m\leq \sum_{k=1}^{K}t_{s,m,k}\log_2(1+\hat{\gamma}_{m,k}), \label{con:ita-sen}\\
~&~\eqref{con:sm-dm-convex}-\eqref{equ:Ek}, \forall\,m \in \mathcal{M}   \text{ and }   k \in \mathcal{K} \label{con:sm-dm-convex2}.
%&\eqref{con:sensing-convex}, \eqref{con:y}, \eqref{con:reporting-convex}.
\end{align}
\end{subequations}
Conventionally, the problem~\eqref{subprob:beamforming} can be solved by using the semi-definite relaxation (SDR) technique, i.e., dropping the rank-one constraint~\eqref{con:rank-w}. As such, it becomes a positive semi-definite program which can be efficiently solved via off-the-shelf optimization tools. However, the solutions $\mathbf{W}_{s,m}$ and $\mathbf{W}_{t,m}$ are not always rank-one.
Instead of the SDR method, similar to~\cite{2020robustBeam}, a penalty function is imposed in the objective~\eqref{obj:beamdorming} to ensure the approximation to the rank-one matrix solutions.

Specifically, the difference between $\mathbf{Tr}(\mathbf{W}_{s,m})$ and $\lambda_{\max}(\mathbf{W}_{s,m})$ is defined as the penalty function, i.e., $\chi_{m}\triangleq\mathbf{Tr}(\mathbf{W}_{s,m}) -  \lambda_{\max}(\mathbf{W}_{s,m})$ and similarly $\tilde{\chi}_{m}\triangleq\mathbf{Tr}(\mathbf{W}_{t,m}) -  \lambda_{\max}(\mathbf{W}_{t,m})$. Adding penalty terms into the objective ~\eqref{obj:beamdorming}, problem~\eqref{subprob:beamforming} can be approximated as follows:
\begin{subequations}\label{subprob:beamforming-penalty}
\begin{align}\label{obj:beamforming-penalty}
\max_{\mathbf{W}_{s,m},\mathbf{W}_{t,m}} ~&~ \hat T(\mathbf{X}(i)) - \kappa_o  \sum_{m=1}^M  \left(\chi_{m}+\tilde{\chi}_{m}\right)\\
 {\rm s.t.}~~~~~&~ \eqref{con:matrix-w}-\eqref{con:sm-dm-convex2},
\end{align}
\end{subequations}
where $\hat T(\mathbf{X}(i))$ is defined by $\sum_{m=1}^{M} \sum_{k=1}^{K}\beta_{m,k}P_m(V+X_k)(a_k+1)$ for notational convenience and $\kappa_o$ is the penalty factor.
The SCA method is further adopted to transform problem~\eqref{subprob:beamforming-penalty} into a convex form in each iteration. Given any feasible $\mathbf{W}_{s,m}^{(\tau)}$ and $\mathbf{W}_{t,m}^{(\tau)}$, the unit eigenvectors $\mathbf{v}^{(\tau)}_{m}$ and $\tilde{\mathbf{v}}^{(\tau)}_{m}$ can be easily determined corresponding to the maximum eigenvalues $\lambda_{\max}(\mathbf{W}_{s,m})$ and $\lambda_{\max}(\mathbf{W}_{t,m})$, respectively.
Then the approximation of the penalty function can be constructed in the $\tau$-th iteration as follows:
\begin{subequations}\label{con:bound-matrix}
\begin{align}
&\chi_{m} = \mathbf{Tr}(\mathbf{W}_{s,m})-(\mathbf{v}^{(\tau)}_{m})^H \mathbf{W}_{s,m} \mathbf{v}^{(\tau)}_{m} ,\\
&\tilde{\chi}_{m} = \mathbf{Tr}(\mathbf{W}_{t,m})-(\tilde{\mathbf{v}}^{(\tau)}_{m})^H \mathbf{W}_{t,m} \tilde{\mathbf{v}}^{(\tau)}_{m} . %\label{con:bound-matrix}
\end{align}
\end{subequations}
The penalty factor $\kappa_o$ can be updated as $\kappa_o^{(\tau+1)} = c\kappa_o^{(\tau)}$ for some positive $c$, if the matrix solution is far from the rank-one approximation. By iteratively updating the penalty factor $\kappa_o$, the objective function can be maximized while ensuring a close approximation to the rank-one solutions~\cite{2020robustBeam}.
%At the convergence, the approximate solution to the problem \eqref{subprob:beamforming-penalty} can be obtained as follows:
%\begin{subequations}\label{con:ws-wr}
%\begin{align}
%\mathbf{w}^{*}_{s,m} = \lambda_{\max}(\mathbf{W}_{s,m})^{1/2}\mathbf{v}_{m},\\
%\mathbf{w}^{*}_{t,m} = \lambda_{\max}(\mathbf{W}_{t,m})^{1/2}\tilde{\mathbf{v}}_{m}.
%\end{align}
%\end{subequations}

The penalty-based iterative method is detailed in Algorithm~\ref{alg-robust-beamforming}. The sensing and forwarding beamforming vectors are initialized to the same vector. In the $\tau$-th iteration, the problem~\eqref{subprob:beamforming-penalty} can be solved efficiently with the matrix solutions $\mathbf{W}_{s,m}^{(\tau)}$ and $\mathbf{W}_{t,m}^{(\tau)}$.
Then, the penalty term $\sum_{m \in \mathcal{M}}\left|\chi_{s,m}^{(\tau)}+\chi_{t,m}^{(\tau)}\right|$ in line 6 of Algorithm~\ref{alg-robust-beamforming} can be evaluated.
If it is greater than the desired accuracy $\Delta$, the penalty factor can be increased as $\kappa_o^{(\tau+1)} = c\kappa_o^{(\tau)}$ in the next iteration. The algorithm will stop until a close approximation of the rank-one solution is found, i.e., the error tolerance is less than $\epsilon$, or the maximum number of iterations is reached, as shown in line $10$ of Algorithm \ref{alg-robust-beamforming}.
%otherwise we end the iteration process and obtain the solution $\mathbf{w}_{s,m}^{*}$ and $\mathbf{w}_{t,m}^{*}$ according to \eqref{con:ws-wr}.

\begin{algorithm}[t]
\caption{The UAVs' Sensing and Transmission Beamforming Optimization Algorithm}\label{alg-robust-beamforming}
	\begin{algorithmic}[1]
		\STATE \textbf{Input:} All GUs' AoI and data states, the UAVs' hovering locations and time allocation strategies.
		\STATE \textbf{Output:} The UAVs' \!beamforming strategy $\{\mathbf{w}_{s,m}, \!\mathbf{w}_{t,m}\}$.
        \STATE \textbf{Initialization:} $\kappa_o^{(0)}$,  $c>1$, $\Delta = 10^{-12}$, $\mathbf{w}_{s,m}^{(0)} = \mathbf{w}_{t,m}^{(0)} = \sqrt{\frac{p_s}{M}}[1,0,...,0]^T$, $\tau_{\max}=15$,  $\tau=1$.
        \REPEAT
        \STATE Solve problem \eqref{subprob:beamforming-penalty} to update $\mathbf{W}_{s,m}^{(\tau)}$ and $\mathbf{W}_{t,m}^{(\tau)}$, and record the objective value as $G^{(\tau)}$.
        \IF{$\sum_{m=1}^{M}\left|\chi_{s,m}^{(\tau)}+\chi_{t,m}^{(\tau)}\right|>\Delta$}
             \STATE  $\kappa_o^{(\tau+1)} \leftarrow c\kappa_o^{(\tau)}$.
        \ENDIF
        \STATE  $\tau \leftarrow  \tau + 1$.
        \UNTIL{$|G^{(\tau+1)}-G^{(\tau)}|\leq \epsilon$ or $\tau>\tau_{\max}$}.
        \STATE Update $\mathbf{w}_{s,m}^{*}$ and $\mathbf{w}_{t,m}^{*}$ by eigenvector decomposition.
	\end{algorithmic}
\end{algorithm}
\subsection{The UAVs' Mobility Control and Time Allocation}
The UAVs' mobility control includes the UAVs' hovering positions and time allocation for the UAVs' flying, sensing, and forwarding phases.
%{\color{blue}We first introduce slack variables $\tilde{\eta}_m$ to simplify the subproblem.}
Given the GUs' access control and the UAVs' beamforming strategies, the UAVs' mobility control can be optimized by solving the following subproblem:
\begin{subequations}\label{subprob:tra-time}
\begin{align}
\max_{\bm{\ell}_m, \bm{t}_m} ~ &\sum_{m=1}^{M}\sum_{k=1}^{K}\beta_{m,k}P_m(V\!\!+\!\!X_k)(a_k+1) \\
 {\rm s.t.}~~~
 &\eqref{con:time-fea}-\eqref{con:trajectory},~ \eqref{equ:sen-time} \text{ and }\eqref{equ:sm-dm}.
\end{align}
\end{subequations}
Problem~\eqref{subprob:tra-time} is difficult to solve due to the non-convex constraints in~\eqref{con:trajectory} and~\eqref{equ:sm-dm}. The squared distance $\|\bm \ell_m-\bm \ell_{\tilde{m}}\|^2$ in~\eqref{con:trajectory} can be approximated by a linear term easily. The squared distance $\|\bm \ell_m-\bm q_{k}\|^2$ also appears in the denominator of the logarithmic function in~\eqref{equ:sm-dm}.
The slack variables $\varphi_m$ can be introduced such that the  constraint~\eqref{equ:sm-dm} can be rewritten as $s_m\leq\varphi_m^2$ and $\varphi_m^2 \leq r_m$.
The first inequality  $s_m\leq\varphi_m^2$ is further transformed into a convex form by the first-order linear approximation as follows:
\begin{align}\label{equ:phi}
s_{m} \leq \left(\varphi_{m}^{(\tau)}\right)^2+2\varphi_{m}^{(\tau)}\Big(\varphi_{m}-\varphi_{m}^{(\tau)}\Big), \quad \forall m \in \mathcal{M}.
\end{align}
Besides, the logarithmic $s_m$  defined in \eqref{equ:ori-sm} can be further approximated is need to be approximated into a simpler form for computational convenience. To this end, introducing the slack variable $z_{m,k}\geq\|\bm \ell_m-\bm q_{k}\|^2$, the received SNR from the GU-$k$ can be reformulated as follows:
\[
\bar{\gamma}_{m,k}=p_s|\Gamma_o|^2 \|\mathbf{g}_{m,k}\|^2\left|\rho \sum_{m' \in \mathcal{M}}\mathbf{g}_{m',k}^H \mathbf{w}_{s,m'}\right|^2 (z_{m',k} z_{m,k})^{-1}.
\]
However, it is still difficult to handle $\bar{\gamma}_{m,k}$ and $\log_2(1+\bar{\gamma}_{m,k})$ directly due to the product $z_{m',k} z_{m,k}$ in the denominator.
The SCA method is resorted to process it iteratively.
\begin{proposition}\label{prop-rs}
Given the feasible solutions $\{z_{m,k}^{(\tau)}\}_{m \in \mathcal{M}, k \in \mathcal{K}}$ in the $\tau$-th iteration of the SCA algorithm, the sensing rate $\log_2(1+\bar{\gamma}_{m,k})$ can be linearly approximated as follows:
\begin{align}\label{equ:G}
& G(z_{m,k}) \triangleq  \log_2\left(1+\bar{\gamma}_{m,k}^{(\tau)}\right) \nonumber \\ &-  \frac{\bar{\gamma}_{m,k}^{(\tau)}}{\ln2 \left(1+\bar{\gamma}_{m,k}^{(\tau)}\right)} \left(\sum_{\tilde{m}=1}^{M}\frac{z_{\tilde{m},k}}{z_{\tilde{m},k}^{(\tau)}}+\frac{z_{m,k}}{z_{m,k}^{(\tau)}} - M-1 \right).
\end{align}
\end{proposition}
The proof of Proposition~\ref{prop-rs} is relegated to Appendix C. Proposition~\ref{prop-rs} gives a linear approximation for the UAV-$m$'s sensing rate $\log_2(1+\bar{\gamma}_{m,k})$. A similar convex reformulation can be applied to $\varphi_m^2\leq r_m$.
In particular, slack variable $\tilde{\bm{\ell}}_m$ can be introduced such that $\tilde{\bm{\ell}}_m\leq \|\bm{\ell}_{m}-\bm{q}_0\|^2$. The received signals at the BS can be reformulated as $\hat{\gamma}_{m,0}= \sum_{m'=m}^{M} p_s\|\mathbf{g}_{m',0}^H\mathbf{w}_{t,m'}\|^2\rho/ \tilde{\bm{\ell}}_{m'}$. Thus, the $\varphi_m^2\leq r_m$ can be reformulated as $\varphi_{m}^2/t_{r,{m}}\leq \log_2(1+\hat{\gamma}_{m,0})-\log_2(1+\hat{\gamma}_{m+1,0})$. By further applying the SCA to the logarithmic function $\log_2(1+\hat{\gamma}_{m,0})$,  $\varphi_m^2\leq r_m$ can be approximated by the following convex form:
\begin{align}\label{equ:reporting-rate}
t^{-1}_{r,{m}}{\varphi_{m}^2}\leq H(\tilde{\bm{\ell}}_m)-\log_2(1+\hat{\gamma}_{m+1,0}),
\end{align}
where $H(\tilde{\bm{\ell}}_m)$ denotes the linear approximation of $\log_2(1+\hat{\gamma}_{m,0})$ and given as follows, similar to that in~\eqref{equ:G}:
\begin{align*}\label{equ:H}
& H(\tilde{\bm{\ell}}_m)  \triangleq  \log_2(1+\hat{\gamma}_{m,0}^{(\tau)}) \nonumber \\ &  - \frac{p_s \rho}{ 1+\hat{\gamma}_{m,0}^{(\tau)}} \sum\limits_{m'=m}^{M} |\mathbf{g}_{m',0}^H\mathbf{w}_{t,m'}|^2\Big(\tilde{\bm{\ell}}_{m'}-\tilde{\bm{\ell}}^{(\tau)}_{m'}\Big)/\Big(\tilde{\bm{\ell}}^{(\tau)}_{m'}\Big)^{2}.
\end{align*}
Till this point, the UAVs' mobility control can be reformulated into the following problem:
\begin{subequations}\label{subprob:convex-tra-time}
\begin{align}
\max_{\bm{\ell}_m, \bm{t}_m} ~ &~\sum_{m=1}^{M}\sum_{k=1}^{K}\beta_{m,k}P_m(V+X_k)(a_k+1) \\
{\rm s.t.}~~ &~ z_{m,k}\geq \|\bm{\ell}_m-\bm{q}_k\|^2 , \\
~ &~s_m\leq \sum_{k=1}^{K}t_{s,m,k}G(z_{m,k}), \\
~ &~\eqref{con:time-fea}-\eqref{con:trajectory},~ \eqref{equ:phi} \text{ and } \eqref{equ:reporting-rate} .
\end{align}
\end{subequations}

The above analysis reveals each step for the AoI-STO algorithm to solve the optimization problem in~\eqref{prob:AoI-lya} following the framework in Fig.~\ref{fig_overall}. It aims to minimize the overall AoI by alternatively optimizing the GUs' access control, the UAVs' mobility control and beamforming strategies, respectively. Given the hovering positions, each UAV first scans the GUs under its coverage and selects a number of GUs according to their current AoI statuses. Then, the UAVs can accept uplink sensing data transmissions from the selected GUs via backscatter communications. To improve the sensing capacity, the UAVs also optimize individuals' beamforming strategies by solving the subproblem in~\eqref{subprob:beamforming-penalty} via an iterative approximation method, as shown in Algorithm~\ref{alg-robust-beamforming}. After that, the UAVs can move to the next sensing positions by trajectory planning. At the end of each FSF time slot, the GUs' AoI queues can be updated according to~\eqref{equ:AoI_queue}.

It is clear that the entire optimization for the original problem in \eqref{prob:AoI-lya} is partitioned into three blocks, i.e., the GUs' access control strategies $\{\beta_{m,k}\}_{m \in \mathcal{M}, k \in \mathcal{K}}$, the UAVs' beamforming strategies $\{\mathbf{w}_{s,m}, \mathbf{w}_{t,m}\}_{m \in \mathcal{M}}$, and the UAVs' mobility control $\{\bm \ell_m, \bm t_m\}_{m \in \mathcal{M}}$.
%It is worth pointing out that in the classical BCD method, the sub-problem for updating each block of variables is required to be solved exactly with optimality in each iteration to guarantee the convergence.
The BCD method for the per-slot control problem in each time slot ends when the error tolerance falls below the threshold $\epsilon$ for two consecutive iterations.
In the evaluation part of Section VI, the threshold $\epsilon$ is set to be $10^{-2}$.
As the objective function is bounded and increasing after each block optimization, the BCD algorithm for the per-slot control problem will finally converge to a stable solution. According to~\cite{2016coverence}, the BCD algorithm has a sub-linear convergence rate and requires $\mathcal{O}(\log(1/\epsilon))$ steps to reach the $\epsilon$-optimal solution. By simulation, it is shown that the overall BCD algorithm for the per-slot control problem in each time slot takes only a few iterations to converge.

Since the GUs' access control is obtained by the heuristic sorting, its complexity is $\mathcal{O}(K^2)$, where $K$ is the number of the GUs.
 %$\mathcal{O}(O_1)$ with $O_1 = K^2$.
The computational complexity of the penalty-based algorithm stems from~\cite{2020robustBeam}.
For a given $\varepsilon>0$, the computational cost for an $\varepsilon$-optimal solution is given by $\mathcal{O}(\ln(1/\varepsilon)\delta\cdot \tau)$,
%in the order of $\ln(1/\varepsilon)\delta$, i.e., $\mathcal{O}(O_2)$ with $O_2 = \ln(1/\varepsilon)\delta\cdot \tau$,
where $\delta$ is the barrier parameter measuring the geometric complexity of the conic constraints.
%For simplicity of analysis of barrier parameter $\delta$, it is assumed that the decision variables $n_d$ in~\eqref{subprob:beamforming-penalty} is real-valued.
The problem in~\eqref{subprob:beamforming-penalty} has $2M+MK$ affine constraints of size $N$ and $2M$ convex constraints. Moreover, the number of decision variables $n_d$ is on the order of $MN^2$. Thus, the barrier parameter $\delta$ is given by $\sqrt{(2M+MK)N+2M}\cdot n_d \cdot \left[(2M+MK)N^3+2M+(2M+MK)N^2+2Mn_d+n_d^2\right] $.
The problem in~\eqref{subprob:convex-tra-time} has been successfully transformed to convex problem using the SCA method and can be effectively addressed.
It involves $5M+MK$ variables constraints, and thus the computational complexity can be approximated by $\mathcal{O}((5M+MK)^{4.5})$~\cite{2022UAVRobust}.
Based on the above analysis, the overall computational complexity of the BCD algorithm for solving the per-slot control problem  is $\mathcal{O}(I(K^2+\ln(1/\varepsilon)\delta\cdot \tau+(5M+MK)^{4.5}))$, where $I$ is the number of iterations.

\begin{table}[t]
	\centering

	\caption{Parameter settings}\label{tab-Sim-Parameters}
    \setlength{\tabcolsep}{3mm}
	\renewcommand\arraystretch{1.2}
    \resizebox{6.5cm}{!}{
	\begin{tabular}{|l|l|}\hline
        \multicolumn{1}{|c|}{\textbf{Parameters}}               &  \multicolumn{1}{c|}{\textbf{Values}} \\\hline
        Number of time slots $T$                                &   100                                \\\hline
		Number of the UAVs $M$                                      &   3                                   \\\hline
		Number of the GUs  $K$                                      &   15                                  \\\hline
        Number of the UAV's antennas  $N$                           &   5                                   \\\hline
        Each GU's uploading mini-slot $t_{s,m,k}$               &   0.2                                \\\hline
       	Reference channel gain $\rho$                           &   $-$30 dB                            \\\hline
		Rician factor $g_0$                                     &   0.94                                \\\hline
		Minimum distance among different UAVs $d_{\rm min}$     &   20 m                                \\\hline
		UAV's maximum flying speed $v_{\max}$                   &   30 m/s                              \\\hline
		UAV's maximum transmit power $p_s$                      &   35 dBm                              \\\hline
		Aggregate noise power $\sigma_u$                        &   $-$110 dBm                           \\\hline
        Upper limit of average AoI  $a_{\max}$       &   15                                  \\\hline
        Control parameter                                       &   $\mu=1$, $V$ = 100                  \\\hline
	\end{tabular}}
\end{table}

\begin{figure}[t]
\centering
\includegraphics[width=0.45\textwidth]{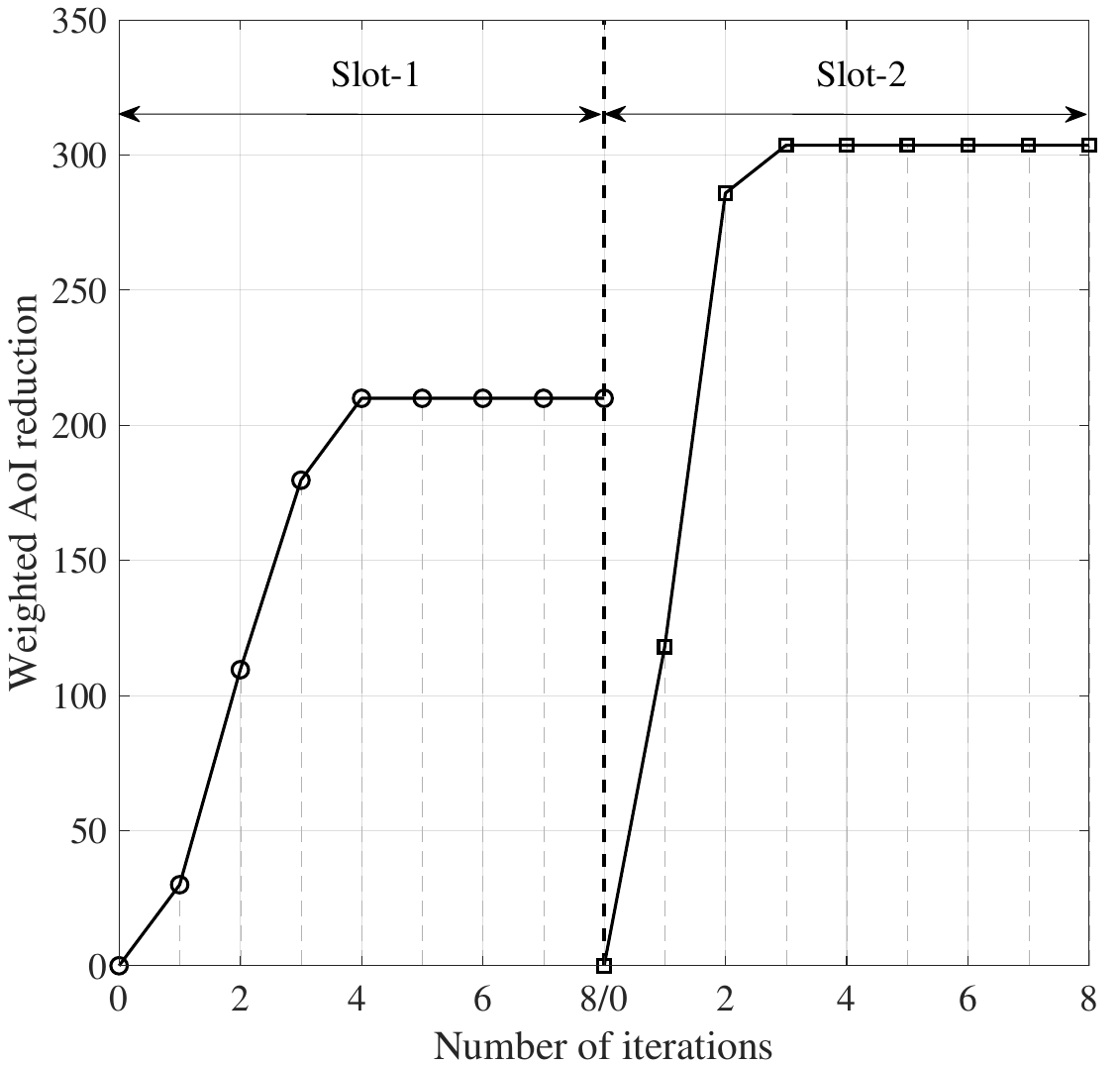}
\caption{Convergence in two time slots of the SCA algorithm}\label{fig:Convergence}
\vspace{-0.4cm}
\end{figure}

\begin{figure}[t]
\centering
\includegraphics[width=0.45\textwidth]{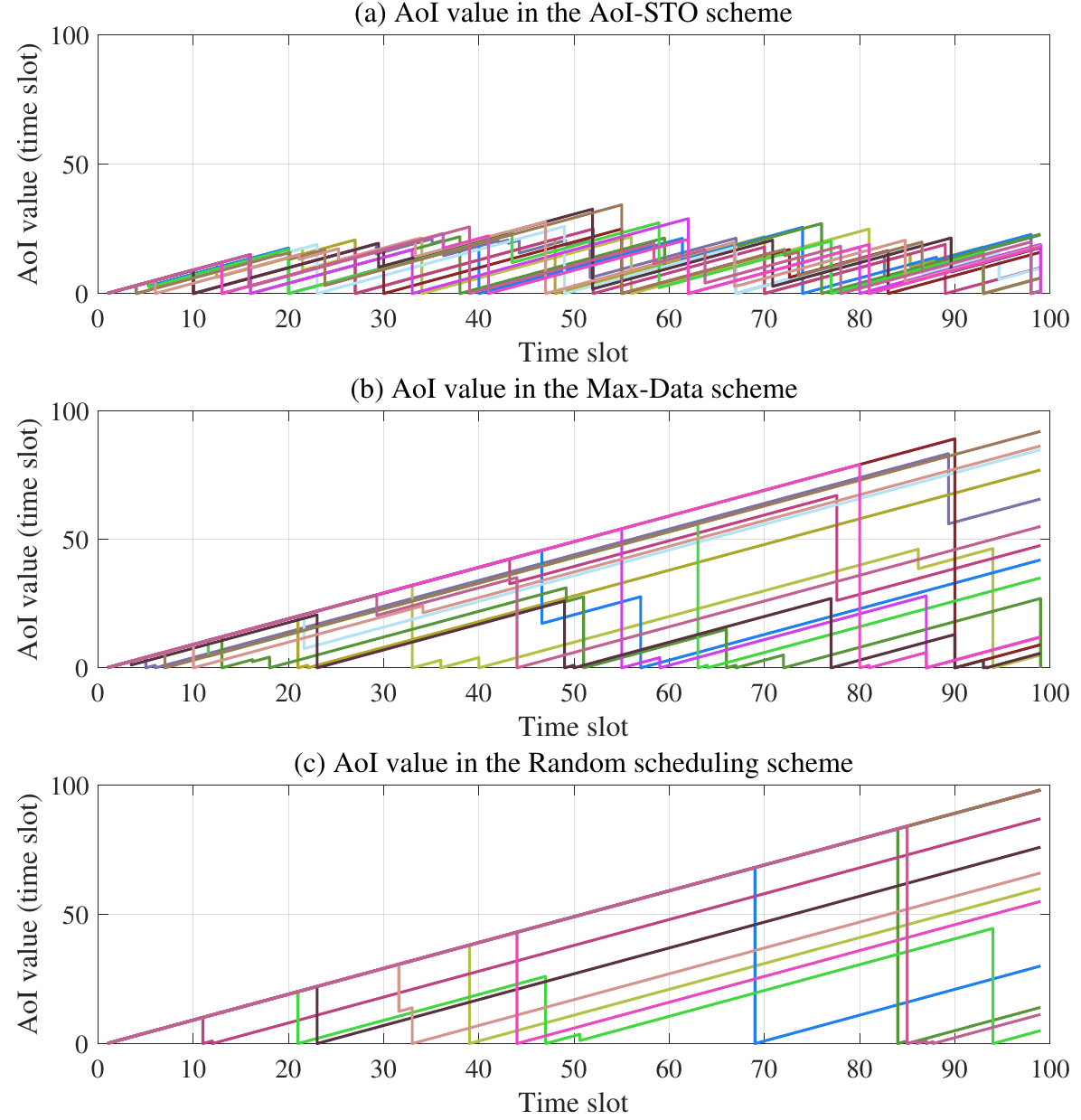}
\caption{AoI dynamics with different access control schemes.}\label{fig:simAOI}\vspace{-0.5cm}
\end{figure}
\section{Numerical Results}\label{sec:eva}
In this section, the performance of the AoI-STO scheme is evaluated. A set of baseline schemes are  also devised for comparison, i.e., the Max-Data scheme and the Random scheduling scheme. The Max-Data scheme means that the GUs with a higher data traffic are first selected to upload their sensing data to the UAVs in each time slot. The Random scheduling scheme allows the GUs to randomly access the uplink GU-UAV channels for data transmission. Without loss of generality, in the simulation $M=3$ UAVs is considered to serve $K=$ 15 GUs in a 500 $\times$ 500 square meter area. The locations of the GUs are randomly distributed. The UAVs' initial coordinates in meters are set as [50, 50, 10], [450, 50, 10], [50, 450, 10], respectively. The BS's location is fixed at [100, 100, 0]. The parameter settings, which are similar to that in~\cite{2018multiUAV-tra-wu} and \cite{2022TMC-AOI}, are summarized in Table~\ref{tab-Sim-Parameters}.

\subsection{AoI Dynamics and Convergence }
Firstly the convergence performance of the SCA algorithm for the per-slot control problem in two consecutive slots is shown in Fig.~\ref{fig:Convergence}. The summation of weighted AoI reduction $\sum_{m=1}^M\sum_{k=1}^{K}\beta_{m,k}P_m(V+X_k)(a_k+1)$, as defined in~\eqref{obj:AoI-lya}, improves significantly within a few iterations and then converges to a stable value, which means that the complexity of the SCA algorithm is affordable. It is observed that the algorithm converges after no more than 5 iterations. Each UAV first searches for a suitable position and then updates a new beamforming strategy. Correspondingly, the weighted AoI reduction first increases slightly due to the change of the UAVs' positions. Then each UAV hovers over the GUs and updates its beamforming strategy to reduce the weighted AoI.

\begin{figure}[t]
\centering
\includegraphics[width=0.45\textwidth]{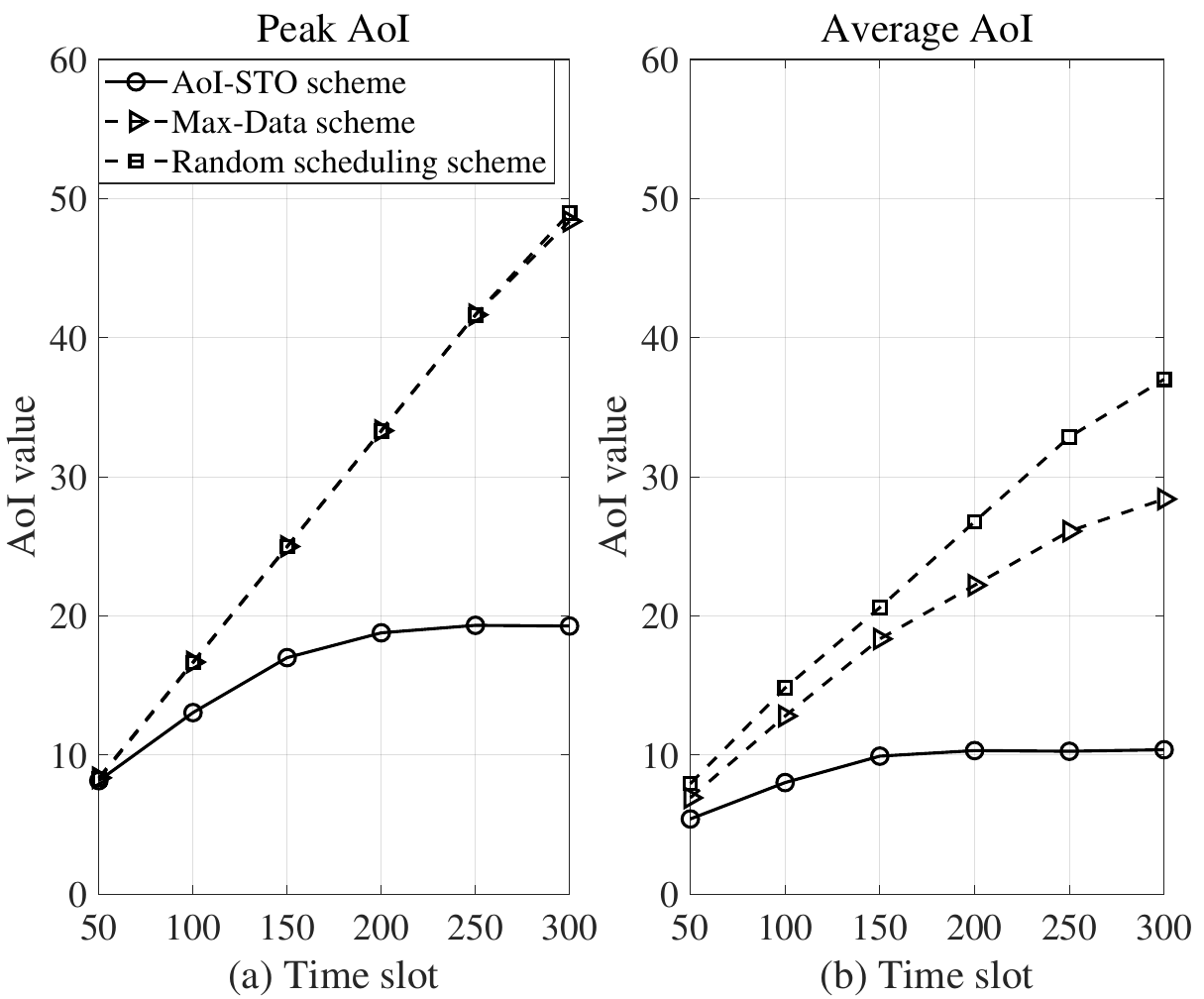}
\caption{Average and peak AoI with different access control schemes}\label{fig:aver-aoi}\vspace{-0.5cm}
\end{figure}

In Fig.~\ref{fig:simAOI}, the real-time dynamics of the AoI performance is plotted in different schemes.
The AoI-STO scheme not only achieves a well-balanced AoI performance, but also guarantees stability of each GU's AoI queue. In the baseline schemes, there exists significant AoI fluctuation among different GUs, i.e., some GUs may have much larger AoI values than that of the other GUs.
This AoI fluctuation is undesirable as it makes the network unstable.
Such fluctuations may be caused by the GUs' access control strategy.
In the AoI-STO scheme, the GUs with a higher AoI and a larger data backlog will be given higher priorities by the UAVs to upload their data. Once the BS successfully receives the sensing data, the GUs complete the update process, and their AoIs can be reduced.
In the Max-Data scheme, the UAVs control the GUs' access according to the data backlogs, which cannot guarantee all GUs' access to the UAVs. Some GUs with small data backlogs may have a larger AoI. The overall AoI will rise up if they cannot access the UAVs in time.
\begin{figure}[t]
\centering
\includegraphics[width=0.45\textwidth]{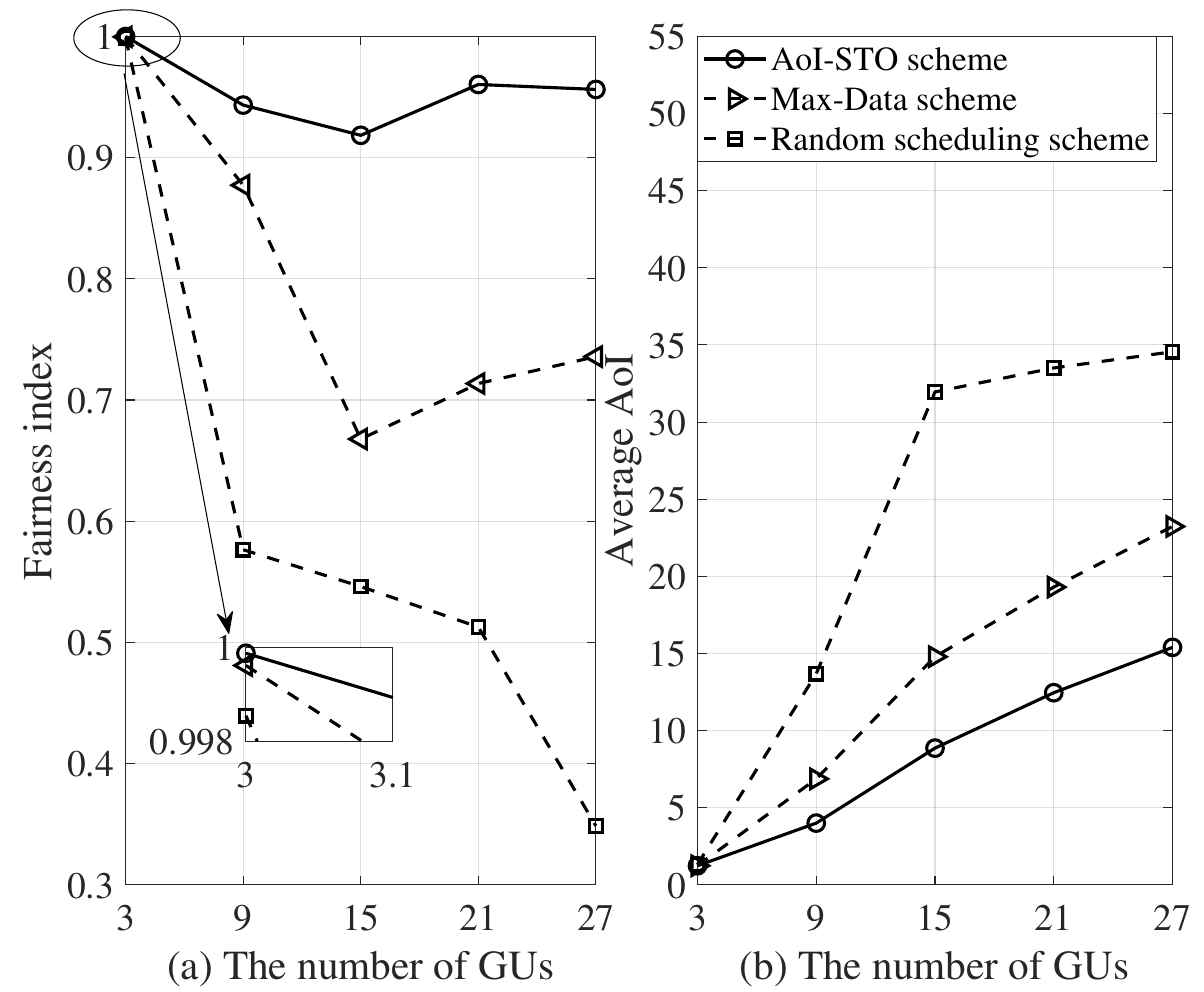}
\caption{The GUs' fairness with different access control schemes.}\label{fig:fairness}
\vspace{-0.4cm}
\end{figure}

A quantitative comparison between different access control schemes is shown in Fig.~\ref{fig:aver-aoi}, where the peak and average AoI values are plotted over different time slots. It is clear that the AoI-STO scheme achieves the lowest peak AoI compared to the baseline schemes.
This is because the proposed GUs' access control can help to adapt the GUs' sensing information uploads to stabilize the GUs' information delay.
Moreover, the AoI-STO scheme can stabilize the peak and average AoI performance, whereas improper access control in the baselines results in a continuous increase of the AoI over time.
The above results indicate that the AoI-STO scheme is suitable for some scenarios with real-time communication requirements.

\subsection{The GUs' Fairness and AoI Performance}
In Fig.~\ref{fig:fairness}, the GUs' fairness is illustrated under different access control schemes.
The GU-$k$'s channel access time in a frame is recorded as $n_k$, and the GUs' fairness can be characterized by the Jain's fairness index $\mathcal{J} = \frac{(\sum_{k=1}^{K} n_k)^2}{K(\sum_{k=1}^{K}n_k^2)}$~\cite{JainFairness},
which ranges from $\frac{1}{N}$ (worst fairness) to $1$ (best fairness). When $J$ becomes large, it implies that different GUs may have comparable channel time for uplink data transmission to upload their sensing data.
%This will lead to a smaller variance in their AoI performance.
%It is observed that the AoI-STO scheme achieves the best fairness index compared with the baseline schemes.
%As the number of GUs increases, the AoI-STO scheme maintains a relatively stable fairness among different GUs, while the fairness index of the baseline schemes  becomes worse off.
This results in a smaller variance in their AoI performance. It is observed that the AoI-STO scheme demonstrates superior fairness performance compared to the baseline schemes. With an increasing number of GUs, the AoI-STO scheme maintains a relatively stable fairness among different GUs.
As the number of GUs grows up,  the GUs' uplink contention becomes severe. In this case, the optimal design of the access control scheme becomes very important to improve the GUs' fairness.
By properly scheduling the GUs' uplink transmission, the GUs' scheduling fairness can be guaranteed in a large-scale UAV-assisted sensing network with massive GUs.
This can prevent channel congestion caused by frequent data uploads and also avoid aging information caused by long waiting delay.
%The comparison in Fig.~\ref{fig:fairness} reveals that the AoI-STO scheme can provide a more stable AoI in terms of fairness by complicated design for the GUs' access control.
\begin{figure}[t]
\centering
\includegraphics[width=0.45\textwidth]{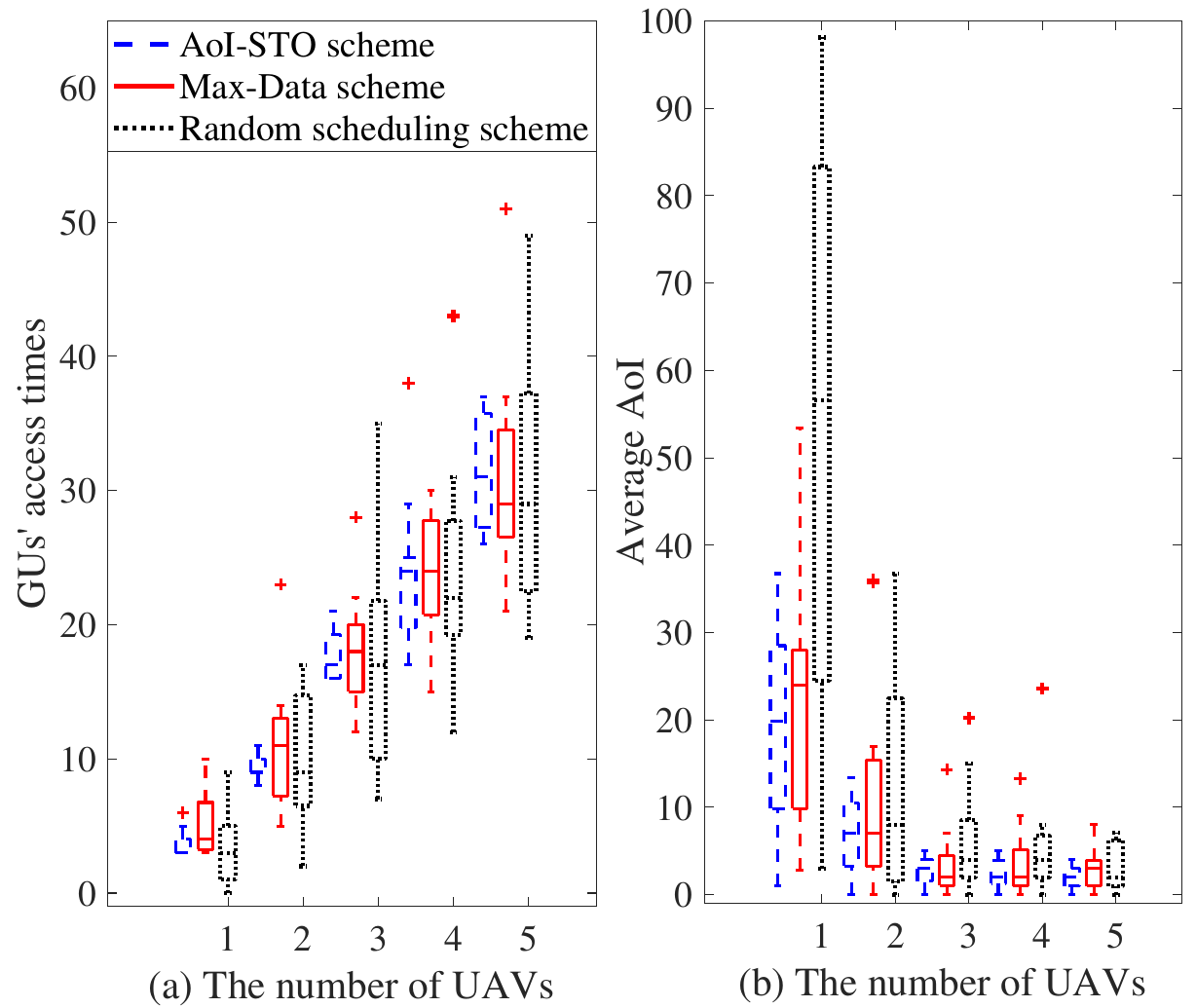}
\caption{The GUs' fairness with different number of UAVs.}\label{fig:fairness-box}
\vspace{-0.4cm}
\end{figure}
Figure~\ref{fig:fairness-box} shows the GUs' access times and average AoI with different access control schemes as the number of UAVs increases.
The box plot of the AoI-STO scheme implies a smaller variance and thus the enhanced fairness among different GUs.
This further confirms the observations in Fig.~\ref{fig:fairness}.
%which corroborates the observations in Fig.~\ref{fig:fairness}.
The long tails in the box plots of the baseline schemes reveal the huge fluctuation in the GUs' AoI performance. As the GUs' instant AoI and data backlogs have been taken into consideration in the AoI-STO scheme, it can minimize the AoI fluctuation comparing to the baseline schemes. A small number of the UAVs becomes difficult to fulfill all GUs' traffic demands. The GUs out of the UAVs' service coverage are hard to connect with the UAVs. Therefore, the reduced access time makes the GUs unable to upload and update data timely, which inevitably leads to an increase in the GUs' AoI values.
This implies that more UAVs can offer the GUs increased  channel access opportunities, which can improve the GUs' sensing capacities and decrease the overall AoI.
%More UAVs can extend the service range and also the channel access opportunities for the GUs. This can increase the GUs' access frequency and decrease the overall AoI.

\subsection{The UAVs' AoI-aware Trajectory Planning}

\begin{figure}[t]
\centering
\subfloat[Case 1]{\includegraphics[width=0.22\textwidth]{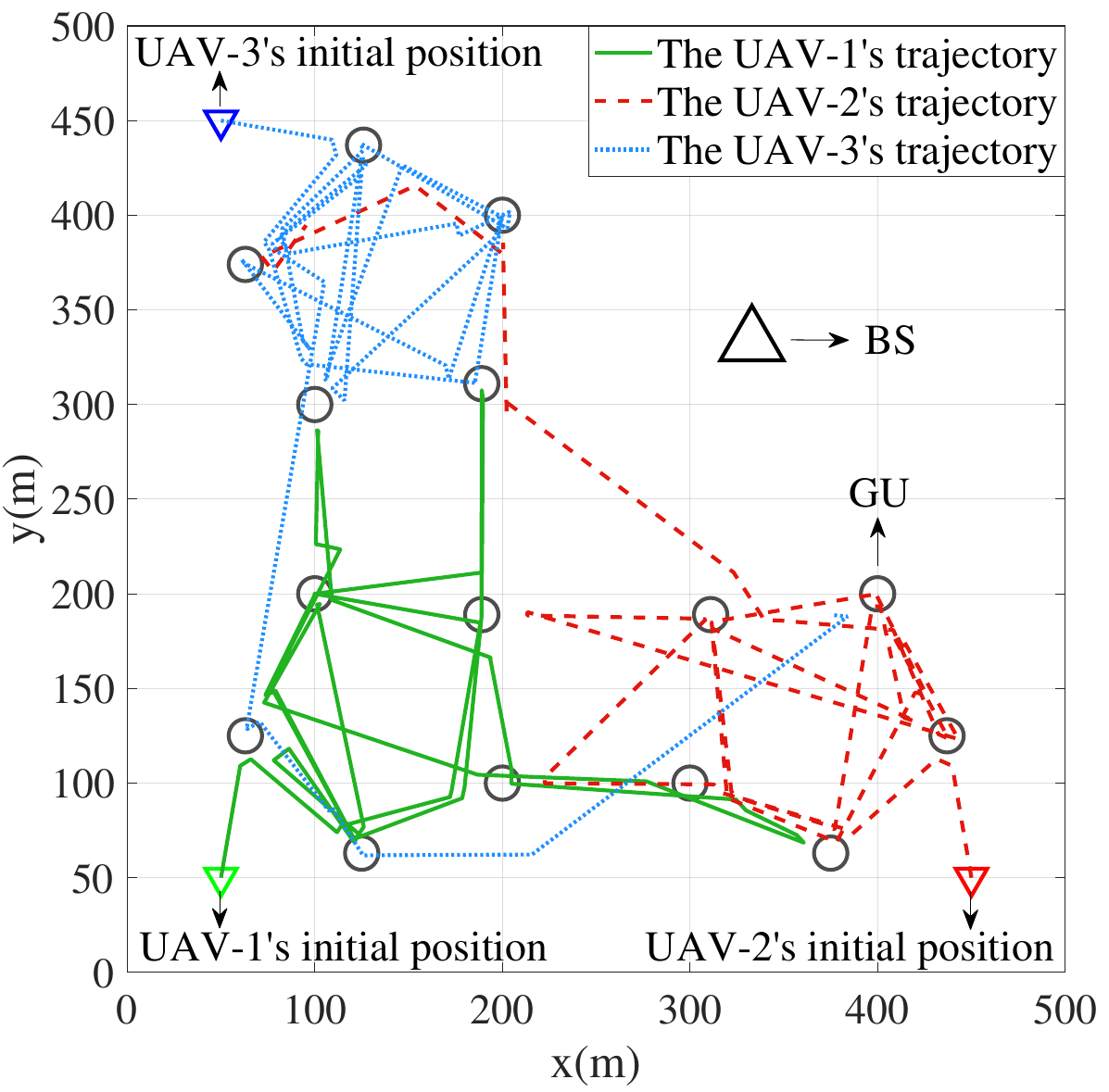}}
\subfloat[Case 2]{\includegraphics[width=0.22\textwidth]{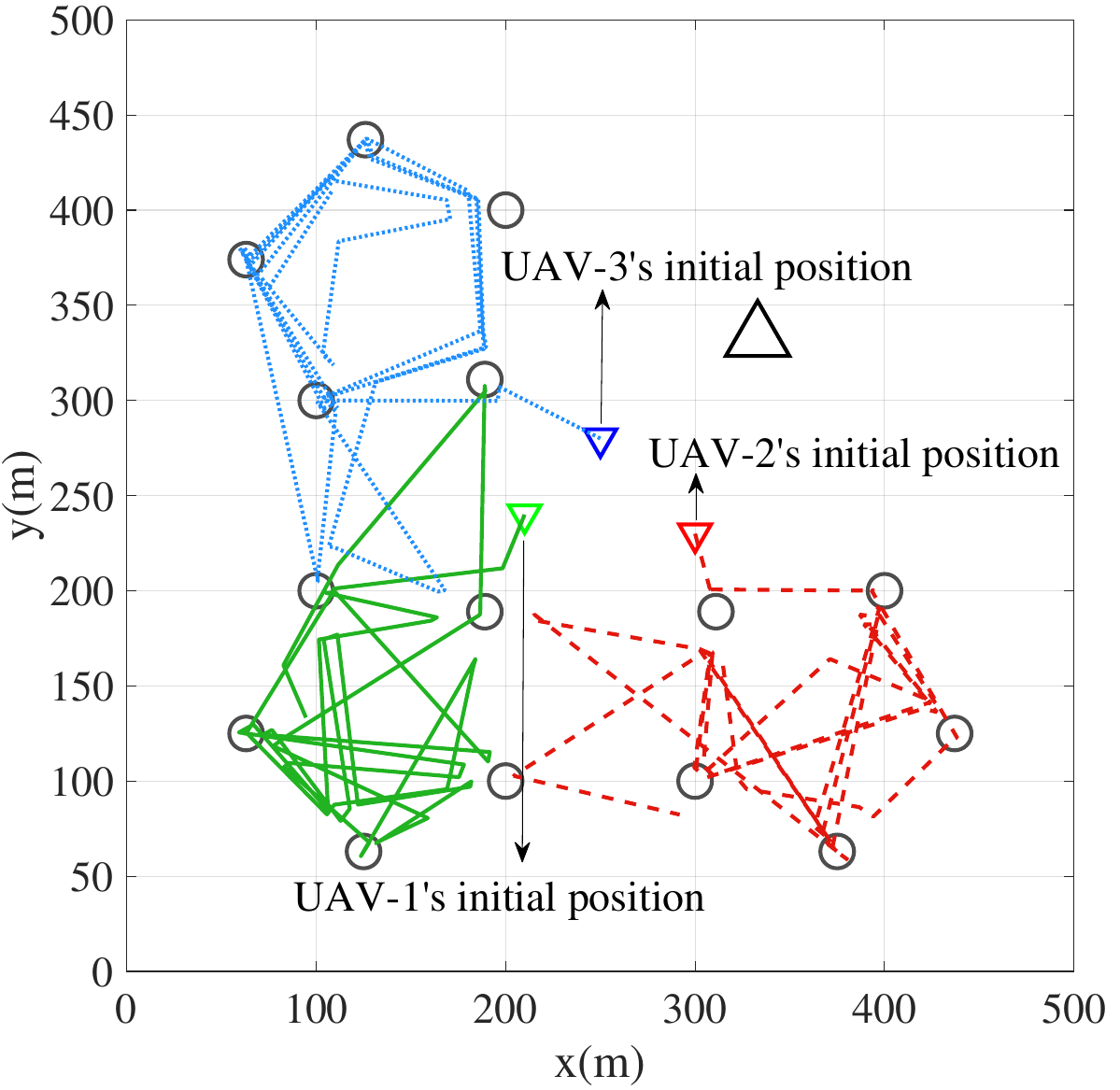}}\\
\subfloat[Case 3]{\includegraphics[width=0.22\textwidth]{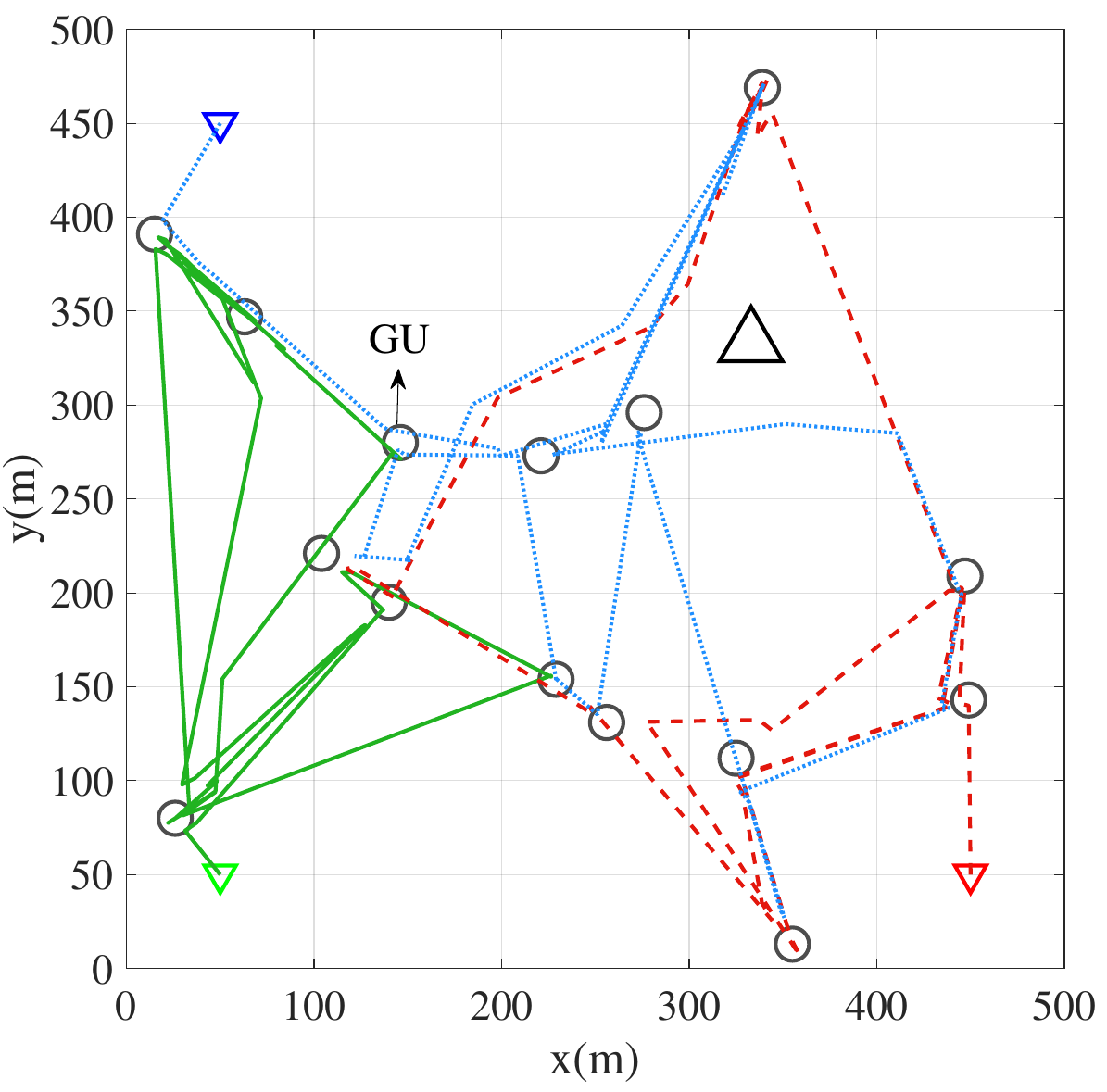}}
\subfloat[Case 4]{\includegraphics[width=0.22\textwidth]{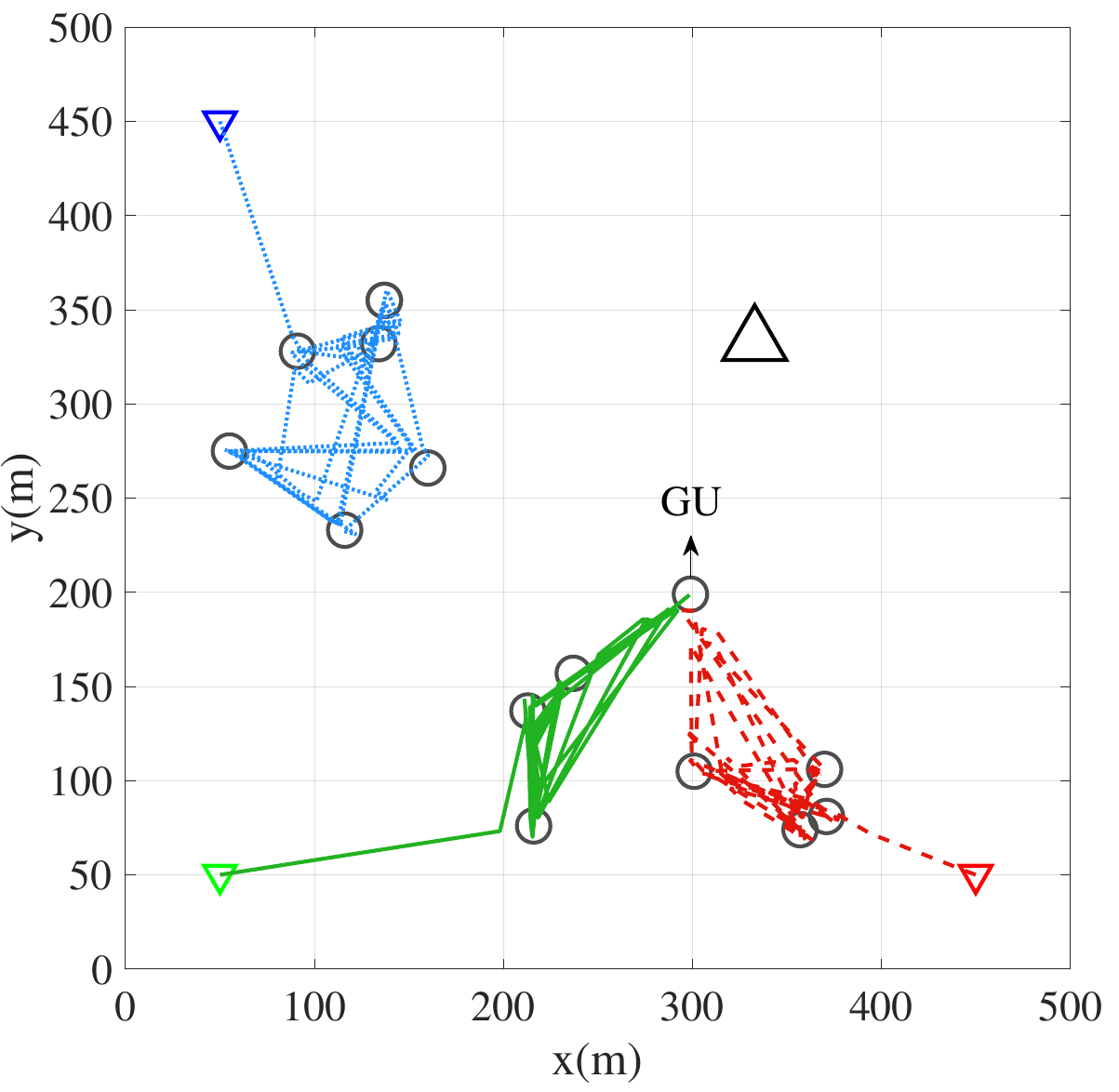}}
\caption{The UAVs' trajectory planning in 4 cases.}\label{fig:trajectory}
\vspace{-0.4cm}
\end{figure}
The UAVs' fast deployment and mobility provide a quick response to the GUs' demands. Specifically, the UAVs can plan their trajectories according to the GUs' spatial distribution and their AoI dynamics.
In this part, the design objective is to show how the UAVs adapt their trajectories from the initial locations according to the GUs' spatial distribution. The following 4 cases are considered in the comparative simulation.
In the cases 1 and 2, the GUs' distribution is the same, while the UAVs' initial positions are different, as shown in Fig.~\ref{fig:trajectory}(a) and Fig.~\ref{fig:trajectory}(b).
In the cases 3 and 4, the UAVs' initial positions are the same, while the GUs' have different spatial distribution, as shown in Fig.~\ref{fig:trajectory}(c) and Fig.~\ref{fig:trajectory}(d).
The UAVs' initial positions in the case 1 are in the corners of the communication area, while the UAVs' initial positions in the case 2 are the same in the center of the service area. The GUs in the case 3 is randomly scattered, while the GUs in the case 4 have a few clustering centers.

\begin{figure}[t]
\centering
\includegraphics[width=0.45\textwidth]{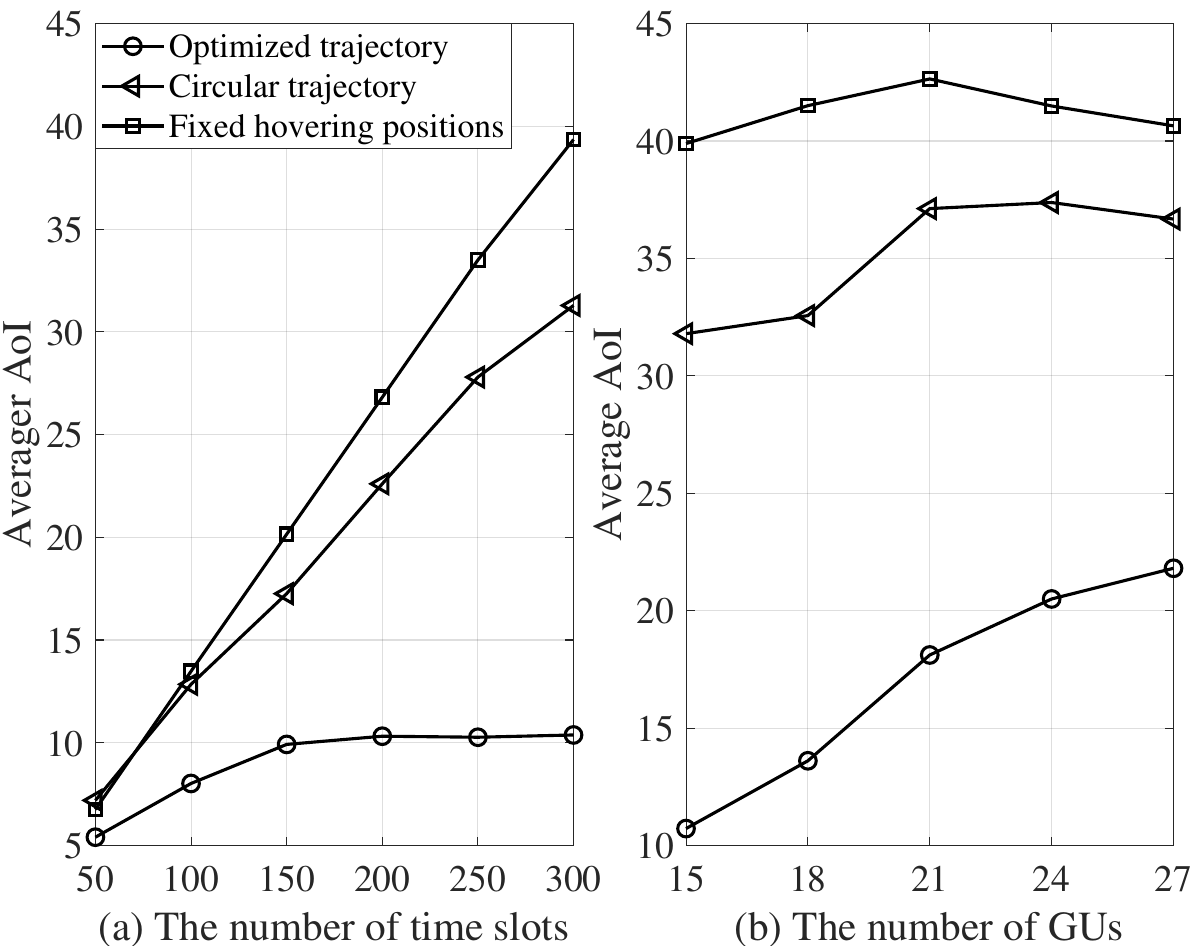}
\caption{Average AoI with different trajectory planning schemes.}\label{fig:aoi_tra}
\vspace{-0.4cm}
\end{figure}
For different cases, the UAVs' trajectories are shown by iteratively solving the optimization problem in~\eqref{prob:AoI-lya}.
It is observed from Fig.~\ref{fig:trajectory} that each UAV can adapt its trajectory to serve different GUs according to their spatial distribution.
Generally, the UAVs' trajectory planning is affected by the GUs' AoI values and locations.
For example, assuming that the GU-$k_1$ is served by the UAV-$m_1$ initially, the GU-$k_1$'s AoI will grow up once the UAV-$m_1$ tends to fly closer to the other GUs and the GU-$k_1$ fails to connect with any UAV.
Then, the GU-$k_1$'s data will become obsolete at the BS due to the untimely scheduling.
Therefore, to avoid continuous growth of the GU-$k_1$'s AoI, the UAV-$m_1$ prefers to fly back to serve the GU-$k_1$ after collecting the other GUs' data.
%All GUs are faced with the same situation.
Assuming that the GUs' traffic is stationary over time, the UAVs prefer to form relatively stable trajectories to serve different GUs.
The UAVs may also deviate from their fixed trajectories when there is a sudden change to the GUs' traffic demands. For example, when the remote GU-$k_2$ has an urgent access demand, the UAV-$m_1$ with the best capability can deviate from its current flying path and then provide the timely service for the GU-$k_2$. Another GU-$k_3$ previously located within the UAV-$m_1$'s service coverage will lose its connection to the UAV-$m_1$. Meanwhile, some other UAVs will fly to serve the GU-$k_3$ and establish a new connection for data collection.

The AoI performance gain caused by the UAVs' trajectory planning is further verified, as shown in Fig.~\ref{fig:aoi_tra}. The GUs' distribution is the same as that of the case 1 in Fig.~\ref{fig:trajectory}(a).
Two trajectory planning schemes are considered for comparison, i.e., the circular trajectory scheme and the fixed hovering scheme. In the circular trajectory scheme, the initial position of each UAV is set at the center of the GUs' cluster, respectively, and each UAV flies with a fixed trajectory that circles around the GUs' cluster. In the fixed hovering scheme, the UAVs are fixed above the center of the GUs' clusters, respectively. It is observed that the overall AoI can be greatly minimized and stabilized by adapting the UAVs' trajectories as shown in Fig.~\ref{fig:aoi_tra}(a). The preset path and fixed positions in baseline schemes both make the UAVs unable to schedule the GUs' data transmissions in time, which results in low information freshness at the BS. The number of the GUs is further increased under different trajectory planning schemes. As the number of the GUs increases, the optimized trajectory can still maintain a low level of AoI by adapting the UAVs' trajectories according to the GUs' distribution. There is a huge gap between the circular trajectory (or the fixed hovering) scheme and the optimized trajectory as shown in Fig.~\ref{fig:aoi_tra}(b), which verifies the superiority of the proposed algorithm.
All the  above results demonstrate that the UAVs can adjust their trajectories based on communication requirements and the GUs' distribution to improve the sensing and transmission capacities in complex and dynamic network environments.

\section{Conclusions}\label{sec:con}
This paper has investigated a multi-UAV-assisted wireless network for minimizing the long-term time-averaged AoI. The AoI minimization problem has been decomposed by the Lyapunov optimization framework. The proposed AoI-STO algorithm can keep information fresh by flexibly optimizing the GUs' access control, the UAVs' mobility, as well as sensing and forwarding beamforming strategies, while maintaining the queue stability. Numerical results have demonstrated that the proposed AoI-STO algorithm can efficiently reduce the overall AoI. In our future work, the UAVs' dynamic altitude control will be introduced to enable more efficient and adaptable navigation in complex network   environments. Moreover, we may focus on designing optimization-driven and distributed learning approach for joint trajectory planning and beamforming optimization to avoid computational demanding operations at the UAVs.
\begin{appendices}
\section{Proof of Proposition \ref{prop-queue}}
%\begin{proposition}\label{prop-queue}
The proof follows a similar idea as that of~\cite{inequality2queue}. If $X_k(i)$ is mean rate stable, i.e., $\lim_{i\rightarrow\infty}\frac{\mathbb{E}\{|X_k(i)|\}}{i}=0$, it ensures the satisfaction of the inequality in \eqref{con:age}.
The queue dynamics in \eqref{equ:AoI_queue} can be relaxed by the following inequality:
\begin{align}
X_k(i+1) &= \big[X_k(i)-a_{\max}\big]^+ + a_k(i+1)\nonumber \\
&\geq X_k(i)-a_{\max} + a_k(i+1).
\end{align}
%and hence we have the following reformulation after simple manipulation:
%\begin{align}
%\frac{X_k(i)-X_k(0)}{i}+\frac{1}{i}\{i\cdot a_{\max}\}\geq \frac{1}{i}\sum_{\tilde{i}=0}^{i-1}a_k(\tilde{i}+1).
%\end{align}
Taking expectation of both sides and inserting $X_k(0)=0$, it easily leads to the inequality $\frac{\mathbb{E}\{X_k(i)\}}{i}+a_{\max}\geq \frac{1}{i}\sum_{\tilde{i}=0}^{i-1}a_k(\tilde{i}+1)$.
%we can build the following inequality:
%\begin{align}
%\frac{\mathbb{E}\{X_k(i)\}}{i}+a_{\max}\geq \frac{1}{i}\sum_{\tilde{i}=0}^{i-1}a_k(\tilde{i}+1).
%\end{align}
Thus, when $\lim_{i\rightarrow\infty}\frac{\mathbb{E}\{|X_k(i)|\}}{i}=0$, , i.e., the virtual queue is mean rate stable, the constraint \eqref{con:age} is satisfied.
\section{Proof of Proposition \ref{prop-bound}}
Noting that $(\max[X-b,0]+a)^2\leq X^2 + a^2 +b^2+2X(a-b)$, the inequality in \eqref{equ:AoI_queue} can be rewritten as $X^2_k(i+1) \leq X^2_k(i) + a^2_k(i+1) + a^2_{\max} + 2X_k(i)(a_k(i+1)-a_{\max})$. After simple manipulation, it results in
$\frac{1}{2}\sum_{k=1}^K \Big(X_k^2(t+1)-X_k^2(i)\Big)
\leq \frac{1}{2}\sum_{k=1}^K \Big(a_k^2(i+1)+a^2_{\max}+2X_k(i)\big(a_k(i+1)-a_{\max}\big)\Big)$. Taking the conditional expectations of both sides yields
\begin{align}\label{con:delta-X}
{\Delta}_L&(\mathbf{X}(i))\leq \frac{1}{2} \sum\nolimits_{k=1}^K \Big(\mathbb{E}\Big[a_k^2(i+1)|\mathbf{X}(i)\Big]+a^2_{\max}\Big)  \nonumber \\
&+\sum\nolimits_{k=1}^K X_k(i)\Big(\mathbb{E}\Big[a_k(i+1)|\mathbf{X}(i)\Big]-a_{\max}\Big).
\end{align}
From the definition of the GUs' AoI in \eqref{equ:aoi-ori}, it is easy to see that $0 \leq a_k(i+1)\leq a_k(i)+1$ and thus $T(\mathbf{X}(i))$ in \eqref{equ:drift-penalty} can be simplified as follows:
\begin{align}\label{con:proof-ddp}
T(\mathbf{X}(i)) \leq \sum_{k=1}^{K}\mathbb{E}\Big[\Big(V+X_k(i)\Big)a_{k}(i+1)\Big|\mathbf{X}(i)\Big] + B,
\end{align}
where $B$ is a finite constant given in~\eqref{equ:Lya_penalty}.
\section{Proof of Proposition \ref{prop-rs}}
The proof of Proposition 3 is straightforward by showing that the sensing rate $\log_2(1+\bar{\gamma}_{m,k})$ is a convex function, and then the linear function can be constructed as its approximation.
To proceed, the convexity of the logarithmic function is proved in the form of $f(x,y)=\log_2\left(\frac{A}{xy}+ B\right )$.
%First, we proof the convexity of function $f(x,y)=\log_2\left(\frac{A}{xy}+ B\right )$.
The Hessian matrix of $f(x,y)$ can be evaluated as follows:
\begin{align*}
\mathbf{H}_f=
\left[\begin{array}{cc}
\frac{A^2+2ABxy}{(Ax+Bx^2y)^2\ln2}&\frac{ABx^2}{(Ax+Bx^2y)^2\ln2}\\\frac{ABy^2}{(Ay+Bxy^2)^2\ln2}&\frac{A^2+2ABxy}{(Ay+Bxy^2)^2\ln2}
\end{array}\right].
\end{align*}
Given $x, y, A, B>0$, it is easy to verify that all elements of the matrix $\mathbf{H}_f$ are positive values. For ease of presentation, $\mathbf{H}_f$ is denoted as $\left[
\begin{array}{cc}
\delta_1&\varsigma_1\\
\varsigma_2&\delta_2
\end{array}\right]$ and it is further verified that $\left|\delta_1\right|>0$ and $\left|\begin{array}{cc}
\delta_1&\varsigma_1\\
\varsigma_2&\delta_2
\end{array}\right|= \delta_1 \delta_2-\varsigma_1\varsigma_2>0$. Hence, the Hessian matrix $H_f$ is positive semi-definite and thus the function $f(x,y)$ is convex. This implies that $\log_2\left(\sum_{j \in \mathcal{J}}\frac{A}{x_jy_j}+ B\right )$ is also a convex function with $A>0, B>0, x_j>0$, and $y_j>0$.
Then, its first-order linear approximation can be easily determined at any feasible point $(x_j^{(\tau)},y_j^{(\tau)})$. As such, the linear approximation in Proposition~\ref{prop-rs} easily follows.
\end{appendices}
\bibliographystyle{IEEEtran}
%\bibliography{your-references}
\bibliography{references}

\begin{thebibliography}{10}
\providecommand{\url}[1]{#1}
\csname url@samestyle\endcsname
\providecommand{\newblock}{\relax}
\providecommand{\bibinfo}[2]{#2}
\providecommand{\BIBentrySTDinterwordspacing}{\spaceskip=0pt\relax}
\providecommand{\BIBentryALTinterwordstretchfactor}{4}
\providecommand{\BIBentryALTinterwordspacing}{\spaceskip=\fontdimen2\font plus
\BIBentryALTinterwordstretchfactor\fontdimen3\font minus \fontdimen4\font\relax}
\providecommand{\BIBforeignlanguage}[2]{{%
\expandafter\ifx\csname l@#1\endcsname\relax
\typeout{** WARNING: IEEEtran.bst: No hyphenation pattern has been}%
\typeout{** loaded for the language `#1'. Using the pattern for}%
\typeout{** the default language instead.}%
\else
\language=\csname l@#1\endcsname
\fi
#2}}
\providecommand{\BIBdecl}{\relax}
\BIBdecl

\bibitem{UAVsurvey}
M.~Mozaffari, W.~Saad, M.~Bennis, Y.-H. Nam, and M.~Debbah, ``A tutorial on {UAVs} for wireless networks: Applications, challenges, and open problems,'' \emph{IEEE Commun. Surv. Tutor.}, vol.~21, no.~3, pp. 2334--2360, Aug. 2019.

\bibitem{2021-Los-Seid}
A.~M. Seid, G.~O. Boateng, B.~Mareri, G.~Sun, and W.~Jiang, ``Multi-agent {DRL} for task offloading and resource allocation in multi-{UAV} enabled {IoT} edge network,'' \emph{IEEE Trans. Netw. Serv. Manag.}, vol.~18, no.~4, pp. 4531--4547, Dec. 2021.

\bibitem{2021UAV-LimitedEnergy}
H.-T. Ye, X.~Kang, J.~Joung, and Y.-C. Liang, ``Optimization for wireless-powered {IoT} networks enabled by an energy-limited {UAV} under practical energy consumption model,'' \emph{IEEE Wireless Commun. Lett.}, vol.~10, no.~3, pp. 567--571, Mar. 2021.

\bibitem{2018UAV-mec-sherman}
N.~Cheng, W.~Xu, W.~Shi, Y.~Zhou, N.~Lu, H.~Zhou, and X.~Shen, ``Air-ground integrated mobile edge networks: {Architecture}, challenges, and opportunities,'' \emph{IEEE Commun. Mag.}, vol.~56, no.~8, pp. 26--32, Aug. 2018.

\bibitem{2022multi-uav-throughput-1}
X.~Liu, B.~Lai, B.~Lin, and V.~C.~M. Leung, ``Joint communication and trajectory optimization for multi-{UAV} enabled mobile internet of vehicles,'' \emph{IEEE Trans. Intell. Transp. Syst.}, pp. 1--13, Jan. 2022, doi:10.1109/TITS.2022.3140357.

\bibitem{2022multi-uav-throughput-2}
W.~Xu, Y.~Sun, R.~Zou, W.~Liang, Q.~Xia, F.~Shan, T.~Wang, X.~Jia, and Z.~Li, ``Throughput maximization of {UAV} networks,'' \emph{IEEE/ACM Trans. Netw.}, vol.~30, no.~2, pp. 881--895, Apr. 2022.

\bibitem{2018shermanA-G-throughput}
S.~Zhang, W.~Quan, J.~Li, W.~Shi, P.~Yang, and X.~Shen, ``Air-ground integrated vehicular network slicing with content pushing and caching,'' \emph{IEEE J. Sel. Areas Commun.}, vol.~36, no.~9, pp. 2114--2127, Sept. 2018.

\bibitem{2022multiUAV-coverage_1}
L.~Wang, H.~Zhang, S.~Guo, and D.~Yuan, ``Deployment and association of multiple {UAVs} in {UAV}-assisted cellular networks with the knowledge of statistical user position,'' \emph{IEEE Trans. Wireless Commun.}, vol.~21, no.~8, pp. 6553--6567, Feb 2022.

\bibitem{2018multiUAV-coverage_2}
L.~Ruan, J.~Wang, J.~Chen, Y.~Xu, Y.~Yang, H.~Jiang, Y.~Zhang, and Y.~Xu, ``Energy-efficient multi-{UAV} coverage deployment in {UAV} networks: {A} game-theoretic framework,'' \emph{China Commun.}, vol.~15, no.~10, pp. 194--209, Oct. 2018.

\bibitem{2018sherman-UAV-deployment}
W.~Shi, J.~Li, W.~Xu, H.~Zhou, N.~Zhang, S.~Zhang, and X.~Shen, ``Multiple drone-cell deployment analyses and optimization in drone assisted radio access networks,'' \emph{IEEE Access}, vol.~6, pp. 12\,518--12\,529, Feb. 2018.

\bibitem{2022multiUAV-aoi-1}
C.~Liu, Y.~Guo, N.~Li, and X.~Song, ``{AoI}-minimal task assignment and trajectory optimization in multi-{UAV}-assisted {IoT} networks,'' \emph{IEEE Internet Things J.}, vol.~9, no.~21, pp. 21\,777--21\,791, Jun. 2022.

\bibitem{2022multiUAV-aoi-2}
R.~Han, Y.~Wen, L.~Bai, J.~Liu, and J.~Choi, ``Age of information aware {UAV} deployment for intelligent transportation systems,'' \emph{IEEE Trans. Intell. Transp. Syst.}, vol.~23, no.~3, pp. 2705--2715, Mar. 2022.

\bibitem{2022YusiUAV-AoI}
Y.~Long, W.~Zhang, S.~Gong, X.~Luo, and D.~Niyato, ``{AoI}-aware scheduling and trajectory optimization for multi-{UAV}-assisted wireless networks,'' in \emph{proc. IEEE GLOBECOM}, Rio de Janeiro, Brazil, Dec. 2022, pp. 2163--2168.

\bibitem{2018UAVair}
Y.~Yang, Z.~Zheng, K.~Bian, L.~Song, and Z.~Han, ``Real-time profiling of fine-grained air quality index distribution using {UAV} sensing,'' \emph{IEEE Internet Things J.}, vol.~5, no.~1, pp. 186--198, Nov. 2018.

\bibitem{2019Disaster-Management}
C.~Luo, W.~Miao, H.~Ullah, S.~McClean, G.~Parr, and G.~Min, \emph{Unmanned Aerial Vehicles for Disaster Management}.\hskip 1em plus 0.5em minus 0.4em\relax Springer Singapore, Aug. 2019.

\bibitem{2017Aoi-origin}
A.~Kosta, N.~Pappas, and V.~Angelakis, ``Age of information: {A} new concept, metric, and tool,'' \emph{Now Foundations and Trends in Netw.}, vol.~12, no.~3, pp. 162--259, Nov. 2017.

\bibitem{2023-AoI-Seid}
A.~M. Seid, A.~Erbad, H.~N. Abishu, A.~Albaseer, M.~Abdallah, and M.~Guizani, ``Multiagent federated reinforcement learning for resource allocation in {UAV}-enabled internet of medical things networks,'' \emph{IEEE Internet Things J.}, vol.~10, no.~22, pp. 19\,695--19\,711, Nov. 2023.

\bibitem{2020uav-sense-trans}
S.~Zhang, H.~Zhang, Z.~Han, H.~V. Poor, and L.~Song, ``Age of information in a cellular internet of {UAVs}: Sensing and communication trade-off design,'' \emph{IEEE Trans. Wireless Commun.}, vol.~19, no.~10, pp. 6578--6592, Jun. 2020.

\bibitem{2018multiUAV-tra-wu}
Q.~Wu, Y.~Zeng, and R.~Zhang, ``Joint trajectory and communication design for multi-{UAV} enabled wireless networks,'' \emph{IEEE Trans. Wireless Commun.}, vol.~17, no.~3, pp. 2109--2121, Jan. 2018.

\bibitem{2022UAV-NOMA-Adam}
A.~B.~M. Adam, M.~S.~A. Muthanna, A.~Muthanna, T.~N. Nguyen, and A.~A.~A. El-Latif, ``Toward smart traffic management with {3D} placement optimization in {UAV}-assisted {NOMA} {IIoT} networks,'' \emph{IEEE Trans. Intell. Transp. Syst.}, pp. 1--11, Jun. 2022, doi:10.1109/TITS.2022.3182651.

\bibitem{2022UAVsensingZhu}
B.~Zhu, E.~Bedeer, H.~H. Nguyen, R.~Barton, and Z.~Gao, ``{UAV} trajectory planning for {AoI}-minimal data collection in {UAV}-aided {IoT} networks by transformer,'' \emph{IEEE Trans. Wireless Commun.}, vol.~22, no.~2, pp. 1343--1358, 2023.

\bibitem{2019sherman-UAB-3DTra}
W.~Shi, J.~Li, N.~Cheng, F.~Lyu, S.~Zhang, H.~Zhou, and X.~Shen, ``Multi-drone {3-D} trajectory planning and scheduling in drone-assisted radio access networks,'' \emph{IEEE Trans. Veh. Technol.}, vol.~68, no.~8, pp. 8145--8158, Aug. 2019.

\bibitem{2022UAVsensingXiong}
X.~Xiong, C.~Sun, W.~Ni, and X.~Wang, ``Three-dimensional trajectory design for unmanned aerial vehicle-based secure and energy-efficient data collection,'' \emph{IEEE Trans. Veh. Technol.}, vol.~72, no.~1, pp. 664--678, Sept. 2022.

\bibitem{2021-UAV-aoi-choudhury}
B.~Choudhury, V.~K. Shah, A.~Ferdowsi, J.~H. Reed, and Y.~T. Hou, ``{AoI}-minimizing scheduling in {UAV}-relayed {IoT} networks,'' in \emph{proc. IEEE 18th International Conference on Mobile Ad Hoc and Smart Systems (MASS)}, Denver, CO, USA, Oct. 2021, pp. 117--126.

\bibitem{2023-UAV-AoI-wang}
X.~Wang, M.~Yi, J.~Liu, Y.~Zhang, M.~Wang, and B.~Bai, ``Cooperative data collection with multiple {UAVs} for information freshness in the internet of things,'' \emph{IEEE Trans. Commun.}, vol.~71, no.~5, pp. 2740--2755, Mar. 2023.

\bibitem{2020-UAV-multiuser-scheduling}
X.~Wu, Z.~Wei, Z.~Cheng, and X.~Zhang, ``Joint optimization of {UAV} trajectory and user scheduling based on {NOMA} technology,'' in \emph{proc. IEEE WCNC}, Virtual Conference, May 2020, pp. 1--6.

\bibitem{2023-UAV-MADDPG}
J.~Du, Z.~Kong, A.~Sun, J.~Kang, D.~Niyato, X.~Chu, and F.~R. Yu, ``{MADDPG}-based joint service placement and task offloading in {MEC} empowered air-ground integrated networks,'' \emph{IEEE Internet Things J.}, pp. 1--1, Oct. 2023, doi:10.1109/JIOT.2023.3326820.

\bibitem{2020-UAV-multiuser-zeng}
H.~Zeng, X.~Zhu, Y.~Jiang, and Z.~Wei, ``{UAV}-ground {BS} coordinated {NOMA} with joint user scheduling, power allocation and trajectory design,'' in \emph{proc. IEEE ICC}, Virtual Conference, Jul. 2020, pp. 1--6.

\bibitem{2021UAV-backscatter-aoi}
X.~Zhang, W.~Luo, Y.~Shen, and S.~Wang, ``Average {AoI} minimization in {UAV}-assisted {IoT} backscatter communication systems with updated information,'' in \emph{2021 IEEE SmartWorld/SCALCOM/UIC/ATC/IOP/SCI}, Virtual Conference, Nov. 2021, pp. 123--130.

\bibitem{2022UAVaoi-cons}
X.~Jia, K.~Zheng, K.~Chi, and X.~Liu, ``{DDPG}-based throughput optimization with {AoI} constraint in ambient backscatter-assisted overlay {CRN},'' \emph{Sensors}, vol.~22, no.~9, pp. 1--20, Apr. 2022.

\bibitem{2021backscatter-delay}
Y.~Ye, L.~Shi, X.~Chu, D.~Li, and G.~Lu, ``Delay minimization in wireless powered mobile edge computing with hybrid backcom and {AT},'' \emph{IEEE Wireless Commun. Lett.}, vol.~10, no.~7, pp. 1532--1536, Apr. 2021.

\bibitem{2022IRS-aoi}
L.~Cui, Y.~Long, D.~T. Hoang, and S.~Gong, ``Hierarchical learning approach for age-of-information minimization in wireless sensor networks,'' in \emph{proc IEEE International Symposium on a World of Wireless, Mobile and Multimedia Networks (WoWMoM)}, Belfast, Northern Ireland, Jun. 2022, pp. 130--136.

\bibitem{2022tvt-uav-irs-aoi}
X.~Fan, M.~Liu, Y.~Chen, S.~Sun, and Z.~Li, ``{RIS}-assisted {UAV} for fresh data collection in {3D} urban environments: {A} deep reinforcement learning approach,'' \emph{IEEE Tran. Veh. Technol.}, vol.~72, no.~1, pp. 632--647, Aug. 2022.

\bibitem{2022-AIRS}
W.~Lyu, Y.~Xiu, S.~Yang, P.~L. Yeoh, Y.~Li, and Z.~Zhang, ``Weighted sum age of information minimization in wireless networks with aerial {IRS},'' \emph{IEEE Trans. Veh. Technol.}, pp. 1--5, Nov. 2022, doi:10.1109/TVT.2022.3223691.

\bibitem{2021UAVIRS-AOI}
M.~Samir, M.~Elhattab, C.~Assi, S.~Sharafeddine, and A.~Ghrayeb, ``Optimizing age of information through aerial reconfigurable intelligent surfaces: A deep reinforcement learning approach,'' \emph{IEEE Trans. Veh. Technol.}, vol.~70, no.~4, pp. 3978--3983, Mar. 2021.

\bibitem{2022channelUAV}
M.~Dai, T.~H. Luan, Z.~Su, N.~Zhang, Q.~Xu, and R.~Li, ``Joint channel allocation and data delivery for {UAV}-assisted cooperative transportation communications in post-disaster networks,'' \emph{IEEE Trans. Intell. Transport. Syst.}, vol.~23, no.~9, pp. 16\,676--16\,689, Aug. 2022.

\bibitem{2021uavTDMA}
B.~Li, R.~Zhang, and L.~Yang, ``Joint user scheduling and {UAV} trajectory optimization for full-duplex {UAV} relaying,'' in \emph{proc. IEEE ICC}, Montreal, Canada, Jun. 2021, pp. 1--6.

\bibitem{2018UAVNOMA-So}
M.~F. Sohail, C.~Y. Leow, and S.~Won, ``Non-orthogonal multiple access for unmanned aerial vehicle assisted communication,'' \emph{IEEE Access}, vol.~6, pp. 22\,716--22\,727, Apr. 2018.

\bibitem{2022UAVPowerAllocation}
M.~Hosseini and R.~Ghazizadeh, ``Stackelberg game-based deployment design and radio resource allocation in coordinated {UAVs}-assisted vehicular communication networks,'' \emph{IEEE Trans. Veh. Technol.}, vol.~72, no.~1, pp. 1196--1210, Sept. 2022.

\bibitem{2022-UAV-NOMA-CR}
V.-H. Dang, L.-M.-D. Nguyen, V.~N. Vo, H.~Tran, T.~D. Ho, C.~So-In, and S.~Sanguanpong, ``Throughput optimization for {NOMA} energy harvesting cognitive radio with multi-{UAV}-assisted relaying under security constraints,'' \emph{IEEE Trans. Cogn. Commun. Netw.}, pp. 1--1, Nov. 2022, doi:10.1109/TCCN.2022.3225165.

\bibitem{2022UAV-NOMA-WBAN}
Z.~Askari, J.~Abouei, M.~Jaseemuddin, A.~Anpalagan, and K.~N. Plataniotis, ``A {Q}-learning approach for real-time {NOMA} scheduling of medical data in {UAV}-aided {WBANs},'' \emph{IEEE Access}, vol.~10, pp. 115\,074--115\,091, Nov. 2022.

\bibitem{2021UAV-backscatter}
P.~X. Nguyen, D.-H. Tran, O.~Onireti, P.~T. Tin, S.~Q. Nguyen, S.~Chatzinotas, and H.~Vincent~Poor, ``Backscatter-assisted data offloading in {OFDMA}-based wireless-powered mobile edge computing for {IoT} networks,'' \emph{IEEE Internet Things J.}, vol.~8, no.~11, pp. 9233--9243, Jan. 2021.

\bibitem{Zhang19TWC_UAV}
G.~Zhang, Q.~Wu, M.~Cui, and R.~Zhang, ``Securing {UAV} communications via joint trajectory and power control,'' \emph{IEEE Trans. Wireless Commun.}, vol.~18, no.~2, pp. 1376--1389, Jan. 2019.

\bibitem{R4}
C.~M.~W. Basnayaka, D.~N.~K. Jayakody, and Z.~Chang, ``Age-of-information-based {URLLC}-enabled {UAV} wireless communications system,'' \emph{IEEE Internet Things J.}, vol.~9, no.~12, pp. 10\,212--10\,223, Jun. 2022.

\bibitem{inequality2queue}
M.~Neely, ``Energy optimal control for time-varying wireless networks,'' \emph{IEEE Trans. Inf. Theory}, vol.~52, no.~7, pp. 2915--2934, Jul. 2006.

\bibitem{2020robustBeam}
Q.~Qi, X.~Chen, and D.~W.~K. Ng, ``Robust beamforming for {NOMA}-based cellular massive {IoT} with {SWIPT},'' \emph{IEEE Tran. Signal Process.}, vol.~68, pp. 211--224, Dec. 2020.

\bibitem{2016coverence}
M.~Hong, M.~Razaviyayn, Z.-Q. Luo, and J.-S. Pang, ``A unified algorithmic framework for block-structured optimization involving big data: {With} applications in machine learning and signal processing,'' \emph{IEEE Signal Process. Mag.}, vol.~33, no.~1, pp. 57--77, Jan. 2016.

\bibitem{2022UAVRobust}
Q.-V. Pham, M.~Le, T.~Huynh-The, Z.~Han, and W.-J. Hwang, ``Energy-efficient federated learning over {UAV}-enabled wireless powered communications,'' \emph{IEEE Trans. Veh. Technol.}, vol.~71, no.~5, pp. 4977--4990, May 2022.

\bibitem{2022TMC-AOI}
M.~Samir, C.~Assi, S.~Sharafeddine, and A.~Ghrayeb, ``Online altitude control and scheduling policy for minimizing {AoI} in {UAV}-assisted {IoT} wireless networks,'' \emph{IEEE Trans. Mob. Comput.}, vol.~21, no.~7, pp. 2493--2505, Jul. 2022.

\bibitem{JainFairness}
A.~B. Sediq, R.~H. Gohary, R.~Schoenen, and H.~Yanikomeroglu, ``Optimal tradeoff between sum-rate efficiency and {Jain}'s fairness index in resource allocation,'' \emph{IEEE Trans. Wireless Commun.}, vol.~12, no.~7, pp. 3496--3509, Jun. 2013.

\end{thebibliography}
\end{document}